\documentclass[aps, prb, reprint, superscriptaddress]{revtex4-1}
\usepackage{graphicx}
\usepackage{dcolumn}
\usepackage{amsmath}
\usepackage{amssymb}
\usepackage{color}
\usepackage{bm}
\usepackage{lipsum} %¿çÀ¸³¤¹«Ê½
\usepackage{braket}
\usepackage{txfonts}
\usepackage[colorlinks=true, letterpaper=true, pdfstartview=FitV, linkcolor=blue, citecolor=blue, urlcolor=blue]{hyperref}

\newcommand{\RNum}[1]{\uppercase\expandafter{\romannumeral #1\relax}}

\newcommand{\1}{\mathbb{I}}
\newcommand{\F}{\mathcal{F}}
\newcommand{\I}{\bm{I}}
\newcommand{\J}{\bm{J}}
\renewcommand{\k}{\bm{k}}
\renewcommand{\u}{\bm{\mu}}
\renewcommand{\r}{\bm{\tilde{r}}}
\renewcommand{\SS}{\bm{\mathcal{S}}_0}
\renewcommand{\S}{\bm{S}}
\renewcommand{\O}{\hat{O}}
\renewcommand{\a}{\bm{a}}
\renewcommand{\b}{\bm{b}}
\renewcommand{\c}{\bm{c}}

\renewcommand{\Re}{\mathrm{Re}}
\newcommand{\Tr}{\mathrm{Tr}}
\renewcommand{\L}{\mathcal{L}}
\renewcommand{\P}{\mathcal{P}}
\newcommand{\PT}{\mathcal{PT}}
\newcommand{\T}{\mathrm{T}}
\newcommand{\B}{\mathcal{B}}
\newcommand{\CC}{\mathcal{C}}
\newcommand{\X}{\mathcal{X}}
\newcommand{\K}{\mathcal{K}}
\newcommand{\N}{\mathcal{N}}
\newcommand{\HP}{\hat{P}}
\newcommand{\vv}{\vec{v}}
\newcommand{\uu}{\vec{u}}

\begin{document}

\title{Dynamical signatures of the Liouvillian Flat Band}

\author{Yu-Guo Liu}
\affiliation{Beijing National Laboratory for Condensed Matter Physics, Institute of Physics, Chinese Academy of Sciences, Beijing 100190, China}

\author{Shu Chen}
\email{schen@iphy.ac.cn}
\affiliation{Beijing National Laboratory for Condensed Matter Physics, Institute of Physics, Chinese Academy of Sciences, Beijing 100190, China}
\affiliation{School of Physical Sciences, University of Chinese Academy of Sciences, Beijing 100049, China}
%\affiliation{Yangtze River Delta Physics Research Center, Liyang, Jiangsu 213300, China}

\date{\today}

\begin{abstract}
Although flat-band structures have been the subject of intensive studies in condensed-matter and optical physics due to their eigenstates, which exhibit huge degeneracy  and allow for the localization of wave packets, it is not clear how the Liouvillian flat band influences the relaxation dynamics of open quantum systems. To this end, we study the dynamical signatures of a Liouvillian flat band in the scheme of a Lindblad master equation. Considering a chain model with gain and loss, we demonstrate three kinds of Liouvillian band dispersion: (i) a flat band, (ii) a dispersionless in the real part, and (iii) a dispersionless band in the imaginary part, and we capture their dynamical signatures. When the Liouvillian rapidity spectrum is flat, the particle numbers in different sites relax to their steady state value with the same decay rate; when the real or imaginary part of rapidity spectrum is dispersionless, the relaxation behaviors have oscillating or forked characteristics. We also show that the Liouvillian flat band can lead to dynamical localization, which is characterized  by a halt in the propagation of local perturbation in the steady state.
\end{abstract}

\maketitle

%%%%%%%%%%%%%%%%%%%%%%%%%%%%%%%%%%%%%%%%%%%%%%%%%%%%%%%%%%%%%%%%%%%%%%%%%%%%%%%%%%%%%%%%%%%%%%%%%%%%%%%%%%%%%%%%%%%%%%%%
%%%%%%%%%%%%%%%%%%%%%%%%%%%%%%%%%%%%%%%%%%%%%%%%%%%%%%%%%%%%%%%%%%%%%%%%%%%%%%%%%%%%%%%%%%%%%%%%%%%%%%%%%%%%%%%%%%%%%%%%
\section{Introduction}
The band structure of a Hamiltonian plays an important role in understanding the motion of particles in periodic crystals. Usually, special band structures may give rise to exotic quantum phenomena. For example, low-energy excitations of electrons on a linear dispersive band in graphene behave like massless Dirac fermions~\cite{Castro2009,Shen2012}. Another instance is the flat band (FB) in which all electrons carry the same energy regardless of their momentum. Due to the dispersionless band structure, particles in a FB have an arbitrarily large effective mass, so they will be localized in real space. Especially, in strongly correlated systems, FB structures indicate a high density of electronic states and the interaction between electrons would become more crucial due to the zero kinetic energy, which leads to rich many-body phenomena.~\cite{Zheng2014,WuCJ,FQH}.

In open quantum systems, the dynamics of density matrix $\rho$ is described by the Lindblad master equation (LME) under the Born-Markov approximation~\cite{Lindblad1976,Gardiner1985,Daniel2020}:
%%%%%%%%%%%%%%%%%%%%%%%%
\begin{equation}\label{Liouvillian}
\frac{d\rho}{dt} = \mathcal{L}(\rho) := -i[H,\rho] + \sum_{\mu} \left( L_{\mu} \rho L_{\mu}^{\dag} - \frac{1}{2} \{ L_{\mu}^{\dag} L_{\mu}, \rho \} \right),
\end{equation}
%%%%%%%%%%%%%%%%%%%%%%%%
where $\mathcal{L}$ is called the Liouvillian superoperator, $H$ is the Hamiltonian of the system, and $L_\mu$ are Lindblad operators which reflect the coupling between system and environment. The Planck constant $\hbar$ is set to unity throughout this paper. There have been several methods developed to obtain the spectrum of $\L$, especially for quadratic systems~\cite{Prosen2008,Chu2017,Naoyuki2019,Horstmann2013,Thomas2022,Yikang2022,Talkington2022}. In Ref. \cite{Talkington2022}, a route for realizing dispersionless bands is proposed based on the underlying mechanism with the emergence of a dissipationless dark space. Generally speaking, the short-time dynamics is related to the Liouvillian eigenvalues with a large modulus of the real part, whereas the long-time relaxation is related to the smallest modulus beyond zero (the so-called Liouvillian gap)~\cite{Cai2013,Marko2015,Mori2020,Haga2021,Nakanishi2022,ZhaiHui2021,Ciuti}. However, how the structure of the Liouvillian, especially the Liouvillian flat band (LFB), influences the dynamics is still a subtle and unexplored question.

In this work, we focus on the dynamics of open quantum systems with LFB. In comparison with the real spectrum of a Hamiltonian system, the Liouvillian spectrum is complex, and thus the corresponding rapidity spectrum can exhibit more rich structures with a dispersionless band in both the imaginary and real parts or either of them. To make our study concrete, we shall first apply a geometrically intuitive method to construct a lattice with correlated gain and loss, which supports LFB, and we explore the generality of dynamical signatures associated with the structure of the Liouvillian spectrum. We show that the rapidity spectra from the Liouvillian and damping-matrix spectra of correlation functions have the same dispersion characteristics, which lead to different signatures of damping dynamics of local particle number distribution: oscillating, forked, synchronous damping are related to the band dispersionless only in imaginary part, real part and in both parts, respectively. Furthermore, we exactly solve the model and show that the LFB can induce dynamical localization, which is characterized by the halt of the propagation of a local perturbation on the non-equilibrium steady state (NESS).

The rest of the paper is organized as follows. In Sec. II,  we first describe the formalism and introduce our model with LFB. In Sec. III, we study the damping dynamics and unveil the dynamical signatures for three kinds of  Liouvillian band dispersion. In Sec. IV, we study compact localized normal master modes, and we discuss the phenomenon of dynamical localization. A summary is given in the final section.

%%%%%%%%%%%%%%%%%%%%%%%%%%%%%%%%%%%%%%%%%%%%%%%%%%%%%%%%%%%%%%%%%%%%%%%%%%%%%%%%%%%%%%%%%%%%%%%%%%%%%%%%%%%%%%%%%%%%%%%%
%%%%%%%%%%%%%%%%%%%%%%%%%%%%%%%%%%%%%%%%%%%%%%%%%%%%%%%%%%%%%%%%%%%%%%%%%%%%%%%%%%%%%%%%%%%%%%%%%%%%%%%%%%%%%%%%%%%%%%%%
\section{Formalism and model}
The density matrix $\rho$ and Liouvillian superoperator $\L$ in Eq.~(\ref{Liouvillian}) can be formally expressed as
%%%%%%%%%%%%%%%%%%%%%%%%
\begin{equation}\label{formal}
\rho=\sum_{\I\J}\rho_{\I\J}|\I\rangle_{\a} \langle\J|_{\a},\ \ \ \L(\rho)=\sum_{ij}\F_i(\a,\a^\dag)\,\rho\,\F_j(\a,\a^\dag),
\end{equation}
%%%%%%%%%%%%%%%%%%%%%%%%
where $\a$ is the set of fermionic annihilation operators i.e. $\a=(a_1,a_2,\cdots)$, $\F_i(\a,\a^\dag)$ is a function with variables among $\a$ and $\a^\dag$, $\I=(I_1,I_2,\cdots)$, $\J=(J_1,J_2,\cdots)$ and
%%%%%%%%%%%%%%%%%%%%%%%%
\begin{equation}
|\I\rangle_{\a} \langle\J|_{\a}=(a^\dag_1)^{I_1}(a^\dag_2)^{I_2}\cdots (a^\dag_L)^{I_L} |0\rangle_{\a} \langle 0|_{\a} (a_L)^{J_L}\cdots (a_1)^{J_1},
\end{equation}
%%%%%%%%%%%%%%%%%%%%%%%%
where $|0\rangle_{\a}$ is the vacuum state for all $a-$fermions. For the convenience of analysis and calculation, we map fermionic LME into a new representation referred to as $\mathcal{C}$ by following the method in Ref.~\cite{Chu2017}:
%%%%%%%%%%%%%%%%%%%%%%%%
\begin{subequations}\label{Mapping}
\begin{align}
&\rho \to |\,\rho\rangle_{\mathcal{C}}=\sum_{\I \J} \rho_{\I \J} (a_1^\dag)^{I_1} \cdots (a_L^\dag)^{I_L} (c_1^\dag \hat{P})^{J_1} \cdots (c_L^\dag \hat{P})^{J_L} \,|0\rangle,~\label{Mrho} \\
&\L \to \hat{L}_{\mathcal{C}}=\sum_{ij}\F_i(\a,\a^\dag) \, \F_j^\T (\hat{P}\c,\c^\dag \hat{P}),~\label{MLZ}
\end{align}
\end{subequations}
%%%%%%%%%%%%%%%%%%%%%%%%
where $\c=(c_1, c_2, \cdots)$ is the set of annihilation operators of $c-$fermions, which is a one-to-one mapping from $\a$, $\T$ means matrix transpose, and $|0\rangle$ is the vacuum state of both $a-$ and $c-$fermions. $\hat{P}$ is the parity operator defined by $\hat{P}=\exp \left( i\pi\sum_{j}(a_j^\dag a_j +c_j^\dag c_j) \right)$, which is introduced to ensure fermionic anticommutation relations between $a-$fermions and $c-$fermions. Full mapping process is shown in Appendix \ref{AP:1}.

\begin{figure}[htbp]\centering
\includegraphics[width=8cm]{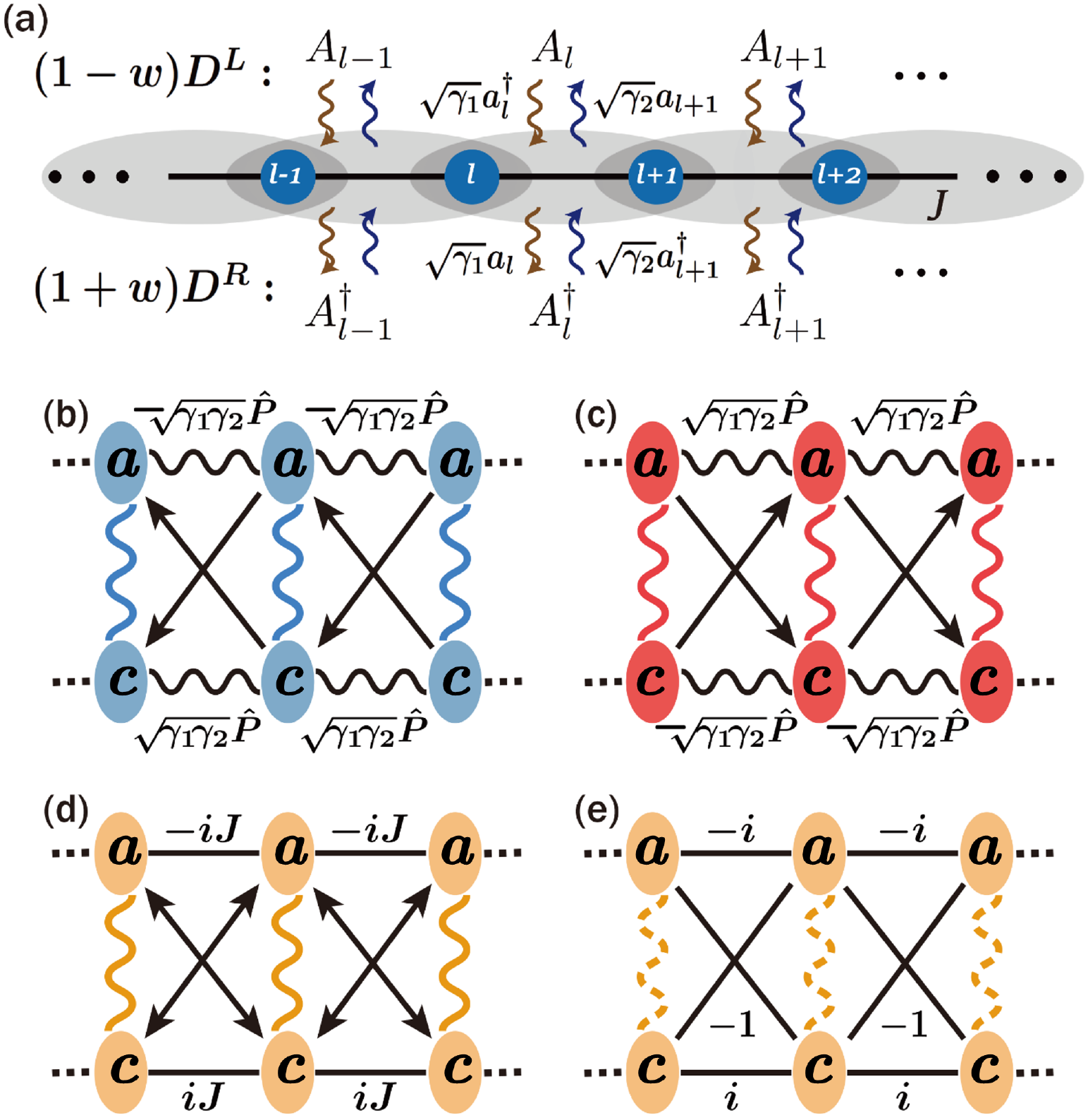}
\caption{ (a) Illustration of the chain hopping $J$ with correlated gain and loss $D^L$ and $D^R$ [Eqs. (5) and (6)]. The environment influences the chain by $D^L$ and $D^R$, which are composed of $A_l$ and $A_l^\dag$. While $A_l$ ties the gain of the $l$ site and loss of the $l+1$ site together, $A_l^\dag$ ties the loss of the $l$ site and gain of the $l+1$ site. The parameter $w$ decides the distribution between $D^L$ and $D^R$. In the representation $\mathcal{C}$ [Eq. (4)], $\hat{D}^L$, $\hat{D}^R$ and $\hat{L}=\hat{H}+\hat{D}^L+\hat{D}^R$ are sketched by (b), (c) and (d). The black arrows indicate directional hoppings with strength $-2\sqrt{\gamma_1 \gamma_2}\hat{P}$. Horizontal black wavy lines represent the pair production and annihilation as $\pm \sqrt{\gamma_1 \gamma_2} \hat{P}(a_l a_{l+1}+h.c.)$ or $\pm \sqrt{\gamma_1 \gamma_2} \hat{P} (c_l c_{l+1}+h.c.)$. The blue, red and orange vertical wavy lines represent $2\gamma_1 \hat{P} a^\dag_l c^\dag_l + 2\gamma_2 \hat{P} c_l a_l$, $2\gamma_2 \hat{P}a^\dag_l c^\dag_l + 2\gamma_1 \hat{P}c_l a_l $ and $2\gamma \hat{P}(a^\dag_l c^\dag_l +c_l a_l) $. The blue, red and orange ovals represent the onsite loss as $(\gamma_1-\gamma_2)\hat{n}_{a/c,\,l} -\gamma_1$, $(\gamma_2-\gamma_1)\hat{n}_{a/c,\,l} -\gamma_2 $ and constant $-\gamma$, where $\hat{n}_{a/c,\,l}$ equals to $a_l^\dag a_l$ or $c_l^\dag c_l$. Part (e) shows (d) in even parity and under flat band condition, where $J=2\sqrt{\gamma_1 \gamma_2}=1 $. The dashed wavy lines indicate that the pairing terms have no effect on single particle- or hole- excitation on its steady state. The solid lines have a cross-stitch-hopping structure, which leads to the flat band by destructive interference.}
\label{fig::1}
\end{figure}

%%%%%%%%%%%%%%%%%%%%%%%%%%%%%%%%%%%%%%%%%%%%%%%%%%%%%%%%%%%%%%%%%%%%%%%%%%%%%%%%%%%%%%%%%%%%%%%%%%%%%%%%%%%%%%%%%%%%%%%%
%%%%%%%%%%%%%%%%%%%%%%%%%%%%%%%%%%%%%%%%%%%%%%%%%%%%%%%%%%%%%%%%%%%%%%%%%%%%%%%%%%%%%%%%%%%%%%%%%%%%%%%%%%%%%%%%%%%%%%%%
%\section{Model}
We consider a Liouvillian in a periodic chain, illustrated in Fig.~\ref{fig::1} (a) :
%%%%%%%%%%%%%%%%%%%%%%%%
\begin{equation}\label{LZ}
\L(\rho)=-i[H,\rho]+(1-w)D^L(\rho)+(1+w)D^R(\rho),
\end{equation}
%%%%%%%%%%%%%%%%%%%%%%%%
where $H=\sum_l J (a_{l+1}^\dag a_l+h.c.)$, $w\in[-1,1]$, and
%%%%%%%%%%%%%%%%%%%%%%%%
\begin{equation}\label{D}
\begin{split}
& D^L(\rho)=\sum_l\left(2 A_l \rho A_l^\dag - A_l^\dag A_l \rho -\rho  A_l^\dag A_l     \right),\\
& D^R(\rho)=\sum_l\left(2 A_l^\dag \rho A_l - A_l A_l^\dag \rho -\rho  A_l A_l^\dag     \right),
\end{split}
\end{equation}
%%%%%%%%%%%%%%%%%%%%%%%%
where $A_l=\sqrt{\gamma_1}a_l^\dag + \sqrt{\gamma_2}a_{l+1}$.  Since $A_l$ and $A_l^\dag$ tie the gain and loss of neighboring sites together,  we can regard that $A_l$ induces a loss of particle-hole pairs (hole on $l$ site and particle on $l+1$ site) and $A_l^\dag$ leads to a gain of the pairs. $D^L$ and $D^R$ represent the influence of the environment. Since they originate from $A_l$ and $A_l^\dag$ respectively, we call $D^L$ and $D^R$ as correlated gain and loss, which could be realized by optical superlattice with a Bose-Einstein condensate reservoir~\cite{Diehl2011}. The role of $w \in [-1,1]$ is analogous to the statistical distribution from temperature~\cite{Landi2022}. When $w>0$ ($w<0$), the gain of particle-hole pairs is stronger (weaker) than the loss.

Mapping Eq.~(\ref{LZ}) into the representation $\CC$, we get a ladder model consisting of an $a-$fermion chain and a $c-$fermion chain. The $\L$ is mapped to
%%%%%%%%%%%%%%%%%%%%%%%%
\begin{equation}
\hat{L}=\hat{H}+(1-w)\hat{D}^L+(1+w)\hat{D}^R,
\end{equation}
%%%%%%%%%%%%%%%%%%%%%%%%
where
%%%%%%%%%%%%%%%%%%%%%%%%
\begin{equation}
\hat{H}=\sum_l \left(-iJ(a_{l+1}^\dag a_l+h.c.)+iJ(c_{l+1}^\dag c_l+h.c.) \right),
\end{equation}
%%%%%%%%%%%%%%%%%%%%%%%%
and the accurate expressions of $\hat{D}^L$ and $\hat{D}^R$ are given in Appendix \ref{AP:2:1}.

$\hat{D}^L$ and $\hat{D}^R$ are illustrated in Figs.~\ref{fig::1} (b) and (c) respectively. $\hat{D}^L$ ($\hat{D}^R$) has the leftward (rightward) cross-stitch-type hoppings along two diagonals of every plaquette in the ladder. As we know, the cross-stitch-type hopping is crucial for generating FB because it can form a destructive-interference structure~\cite{Creutz1999,Maimaiti2017,Maimaiti2021,Kuno2020}. In our model, no matter how to distribute the proportion of the correlated gain and loss $\hat{D}^L$ and $\hat{D}^R$ by $w$, the cross-stitch hopping always exists in representation $\CC$, which is the origin of our FB.
%%%%%%%%%%%%%%%%%%%%%%%%
%%%%%%%%%%%%%%%%%%%%%%%%

\begin{figure}[htbp]\centering
\includegraphics[width=8.5cm]{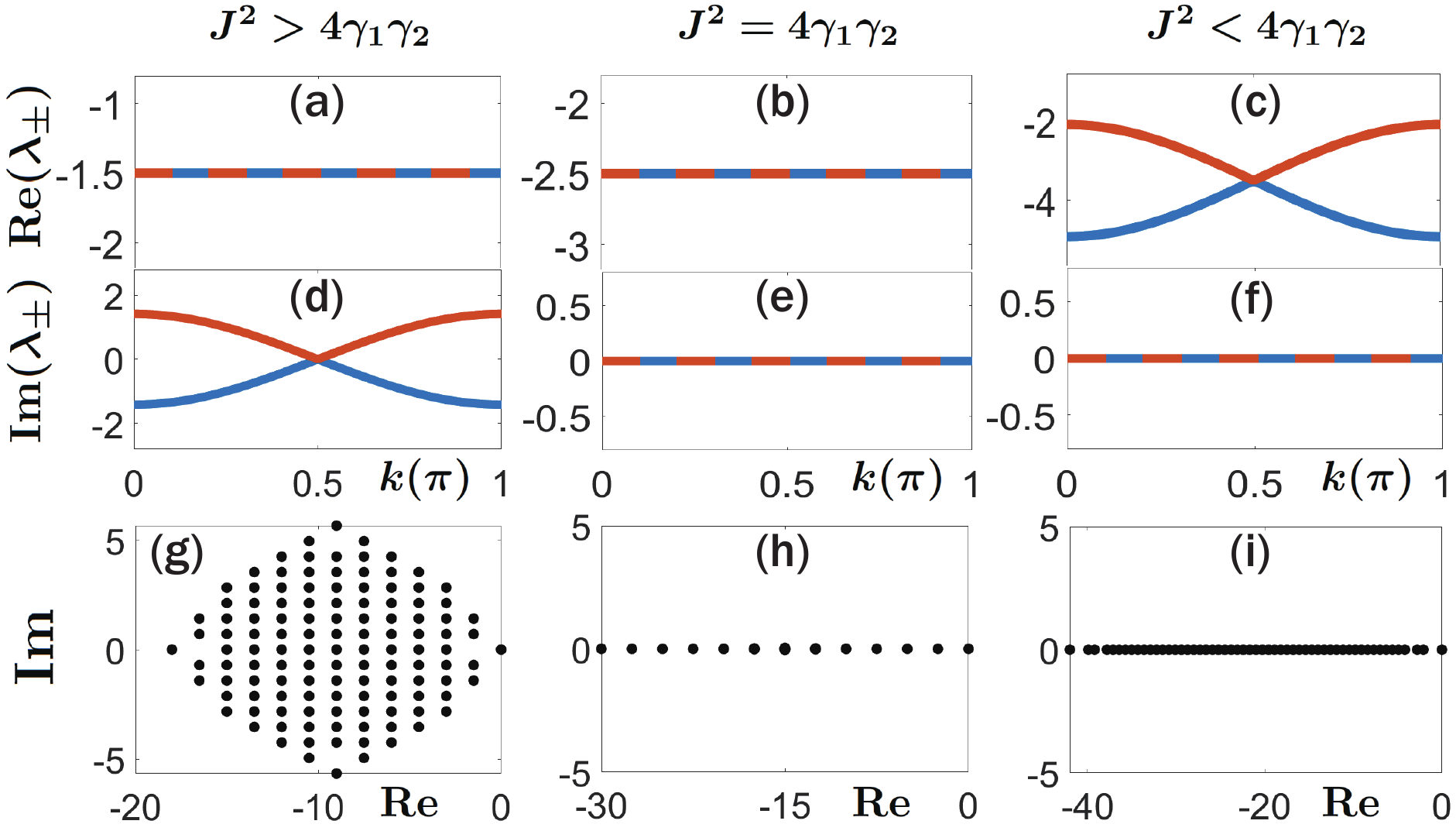}
\caption{(a)$\sim$(c) the real part of rapidity spectra $\lambda_{\pm}(k)$. (d)$\sim$(f) the imaginary part of $\lambda_{\pm}(k)$. $\lambda_{\pm}(k)$ is independent with $w$. The Liouvillian spectra can be constructed by summing different numbers of rapidities. Here, we show the Liouvillian spectra obtained by exactly diagonalizing 6-site lattice in (g)$\sim$(i) with $w=0$, $J=1$ and $\gamma_1=0.25$. $\gamma_2=0.5$ in (a), (d) and (g). $\gamma_2=1$ in (b), (e) and (h). $\gamma_2=1.5$ in (c), (f) and (i).}
\label{fig::2}
\end{figure}

In momentum space, $\hat{L}$ can be expressed in BdG form as
%%%%%%%%%%%%%%%%%%%%%%%%
\begin{eqnarray}
& \hat{L}=0.5\,\hat{L}_{k=0}+\sum_{k=0^+}^{\pi^-} \hat{L}_k,\label{Lk0}\\
& \hat{L}_k=(a^\dag_k \ c^\dag_k \ a_{-k} \ c_{-k})\ \L_k \ (a_k \ c_k \ a^\dag_{-k} \ c^\dag_{-k} )^{\T} -4\gamma,\label{Lk05}
\end{eqnarray}
%%%%%%%%%%%%%%%%%%%%%%%%%
where $\gamma=\gamma_1+\gamma_2$. Due to parity conservation in $\hat{L}$, the operator $\hat{P}$ can be substituted by a constant $P$ which equals $1$ or $-1$ when $\hat{L}$ acts on the state with even or odd fermions. Then we have
%%%%%%%%%%%%%%%%%%%%%%%%
\begin{eqnarray}\label{Lk1}
&\L_k=-i2J\cos{k} \sigma_z \otimes \sigma_z -4\sqrt{\gamma_1 \gamma_2} \cos{k} P \sigma_z \otimes \sigma_x  - \nonumber\\
&2\gamma P \sigma_y \otimes \sigma_y \nonumber   +   2w\left[
 (\gamma_2-\gamma_1)\sigma_z \otimes \1 +2\sqrt{\gamma_1\gamma_2}\sin{k} \sigma_y \otimes \sigma_z \right. \nonumber\\
& \left. +i (\gamma_2 -\gamma_1)P\sigma_x \otimes \sigma_y +i 2\sqrt{\gamma_1\gamma_2}\sin{k} P \1 \otimes \sigma_x
 \right],
\end{eqnarray}
%%%%%%%%%%%%%%%%%%%%%%%%
where $\1$ and $\sigma_i$ are identity and Pauli matrices. $\hat{L}_k$ can be diagonalized as
%%%%%%%%%%%%%%%%%%%%%%%%
\begin{equation}\label{Lk+-}
\begin{split}
& \hat{L}_k=\lambda_{-}(k) \left(\zeta^{'}_1(k) \zeta_1(k)+\zeta^{'}_4(k) \zeta_4(k)\right)\\
&+\lambda_{+}(k) \left( \zeta^{'}_2(k) \zeta_2(k)+\zeta^{'}_3(k) \zeta_3(k) \right),
\end{split}
\end{equation}
%%%%%%%%%%%%%%%%%%%%%%%%
where $\zeta^{'}_i(k)$ and $\zeta_j(k^{'})$ fulfill anticommutation relations: $\{\zeta^{'}_i (k),\zeta_j (k^{'})\}=\delta_{ij}\delta_{kk^{'}}$ and $\{\zeta^{'}_i(k),\zeta^{'}_j(k^{'})\} = \{\zeta_i(k),\zeta_j(k^{'})\}=0$. The $\lambda_{\pm}(k)$ is called the rapidity spectrum~\cite{Prosen2008}. In this model, $\lambda_{\pm}(k)$ has the same expression in both odd and even parity~\cite{Exp1}:
%%%%%%%%%%%%%%%%%%%%%%%%
\begin{equation}
\lambda_{\pm}(k)=-2\gamma \pm 2 m_k,
\end{equation}
%%%%%%%%%%%%%%%%%%%%%%%
where
%%%%%%%%%%%%%%%%%%%%%%%%
\begin{equation} \label{mk}
m_k=\left\{\begin{matrix} \sqrt{(4\gamma_1\gamma_2-J^2)\cos^2 k}, \ \ \ J^2 \le 4\gamma_1\gamma_2, \\ i\sqrt{(J^2-4\gamma_1\gamma_2)\cos^2 k}, \ \ \ J^2 > 4\gamma_1\gamma_2.  \end{matrix} \right.
\end{equation}
%%%%%%%%%%%%%%%%%%%%%%%%
As shown in Figs.~\ref{fig::2} (a)$\sim$(f), when $J^2=4\gamma_1 \gamma_2$, $\lambda$ is a FB of $k$. When $J^2<4 \gamma_1 \gamma_2$ ($J^2>4 \gamma_1 \gamma_2 $), $\lambda$ is dispersionless in its imaginary (real) part.
%%%%%%%%%%%%%%%%%%%%%%%%
%%%%%%%%%%%%%%%%%%%%%%%%

The rapidity spectrum corresponds to a single-mode decaying. The Liouvillian spectrum is constructed by summing different numbers of rapidities. Therefore, the Liouvillian spectrum inherits the characteristics of a rapidity spectrum, as shown in Figs.~\ref{fig::2} (g)$\sim$(i). When $J^2=4 \gamma_1 \gamma_2$, the Liouvillian spectrum only consists of some highly degenerate discrete points [Fig.~\ref{fig::2} (h)], corresponding to different numbers of occupations of the flat-band rapidity spectrum, so we call this kind of Liouvillian spectrum as the LFB.

%%%%%%%%%%%%%%%%%%%%%%%%%%%%%%%%%%%%%%%%%%%%%%%%%%%%%%%%%%%%%%%%%%%%%%%%%%%%%%%%%%%%%%%%%%%%%%%%%%%%%%%%%%%%%%%%%%%%%%%%
%%%%%%%%%%%%%%%%%%%%%%%%%%%%%%%%%%%%%%%%%%%%%%%%%%%%%%%%%%%%%%%%%%%%%%%%%%%%%%%%%%%%%%%%%%%%%%%%%%%%%%%%%%%%%%%%%%%%%%%
%\section{The symmetry and exceptional point of Liouvillian}
Due to $\hat{L}_k=\hat{L}_{-k}$, Eq.~(\ref{Lk0}) can be rewritten as $\hat{L}=\frac{1}{2} \sum_{k=-\pi}^\pi \hat{L}_k$. With $k \in (-\pi, \pi)$, it is easy to check that $\L_k$ in Eq.~(\ref{Lk1}) has time-reversal symmetry (TRS), particle-hole symmetry (PHS) and chiral symmetry (CS)\cite{Kawabata2019,Ludwig2015,ChunHui2019a,ChunHui2019b}:
%%%%%%%%%%%%%%%%%%%%%%%%
\begin{equation}~\label{SofL}
\begin{split}
&\mathrm{TRS}:\ \mathcal{T}_{+}\  \L^{*}_k\  \mathcal{T}_{+}^{-1} = \L_{-k}\ \ \Longrightarrow \ \ \mathcal{T}_{+}=\sigma_z \otimes \sigma_x; \ \mathcal{T}_{+} \mathcal{T}_{+}^{*}=1 \\
&\mathrm{PHS}:\ \mathcal{C}_{-}\  \L^{\T}_k\  \mathcal{C}_{-}^{-1} = -\L_{-k}\ \ \Longrightarrow \ \ \mathcal{C}_{-}=\sigma_x \otimes \1; \ \mathcal{C}_{-} \mathcal{C}_{-}^{*}=1 \\
&\mathrm{CS}:\ \Gamma\  \L^{\dag}_k\  \Gamma^{-1} = -\L_{k}\ \ \Longrightarrow \ \ \Gamma =\sigma_y \otimes \sigma_x; \ \Gamma^2=1.
\end{split}
\end{equation}
%%%%%%%%%%%%%%%%%%%%%%%
Due to the fact that $\L_k$ has a full pure real spectrum in the region $J^2<4\gamma_1 \gamma_2$ as shown in Figs.~\ref{fig::2} (c) and (f), the mathematical theorem ensures that the Liouvillian has pseudo-Hermiticity~\cite{Mostafazadeh2002a,Mostafazadeh2002b,Yuto2020} i.e. there exists a Hermitian matrix $\eta$ in which $\eta\ \L^{\dag}_k \eta^{-1}=\L_k $. In addition, the complex spectrum in Figs.~\ref{fig::2} (a) and (d) shows the breaking of pseudo-Hermiticity. Especially, when $w=0$, the system will additionally have inversion symmetry (IS) and the pseudo-Hermiticity will be enhanced to the parity-time symmetry (PTS):
%%%%%%%%%%%%%%%%%%%%%%%%
\begin{equation}
\begin{split}
&\mathrm{IS}:\ \P \ \L_k \ \P^{-1} = \L_{-k}\ \ \Longrightarrow \ \ \P=\1 \otimes \1 \\
&\mathrm{PTS}:\ \PT \ \L_k^{*}\ \PT^{-1} = \L_k\ \ \Longrightarrow \ \ \PT=\sigma_x \otimes \sigma_z.
\end{split}
\end{equation}
%%%%%%%%%%%%%%%%%%%%%%%

\begin{figure}[htbp]\centering
\includegraphics[width=8.5cm]{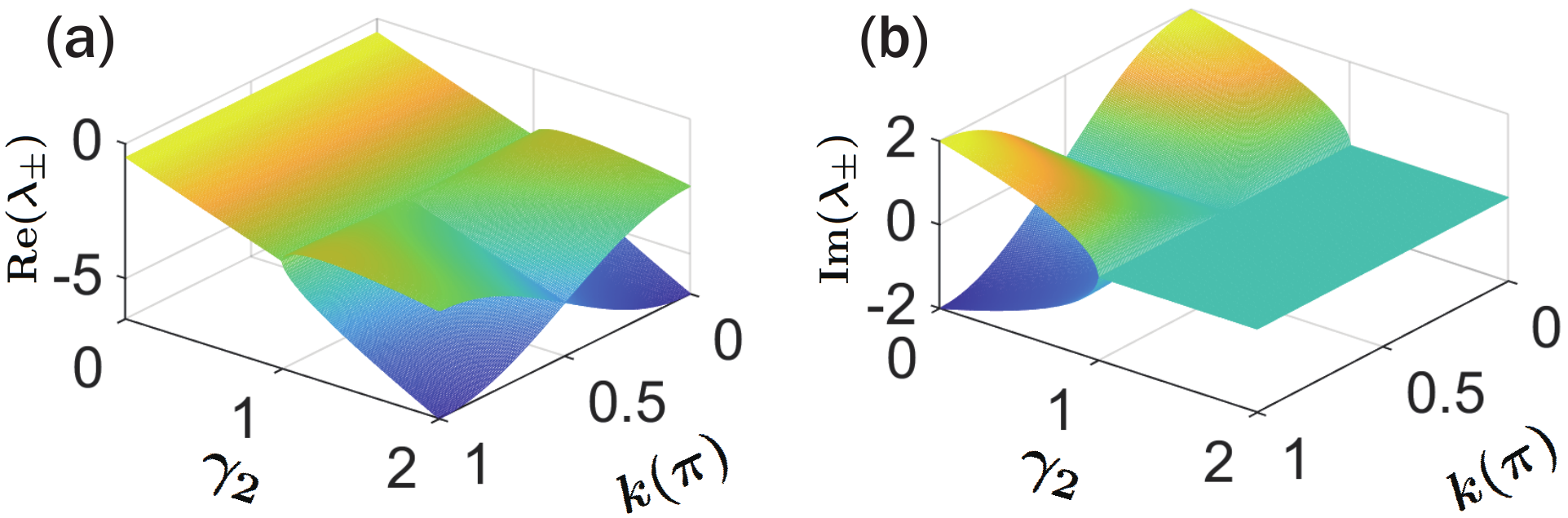}
\caption{ The real (a) and imaginary (b) part of $\lambda_{\pm} (k)$ as a function with $k$ and $\gamma_2$. Other parameters are taken as $J=1$ and $\gamma_1=0.25$ }
\label{fig::3}
\end{figure}

When $J^2=4\gamma_1 \gamma_2 $, the exceptional point of $\L_k$ emerges. To see it clearly, we show the real and imaginary part of the rapidity $\lambda_{\pm} (k)$ in Fig.~\ref{fig::3}. When the flat-band condition is satisfied ($\gamma_2=1$), exceptional degeneracy occurs between $\lambda_+$ and $\lambda_-$.
%%%%%%%%%%%%%%%%%%%%%%%%
%%%%%%%%%%%%%%%%%%%%%%%%
%%%%%%%%%%%%%%%%%%%%%%%%%%%%%%%%%%%%%%%%%%%%%%%%%%%%%%%%%%%%%%%%%%%%%%%%%%%%%%%%%%%%%%%%%%%%%%%%%%%%%%%%%%%%%%%%%%%%%%%%
%%%%%%%%%%%%%%%%%%%%%%%%%%%%%%%%%%%%%%%%%%%%%%%%%%%%%%%%%%%%%%%%%%%%%%%%%%%%%%%%%%%%%%%%%%%%%%%%%%%%%%%%%%%%%%%%%%%%%%%%
\section{Damping dynamics}
%\section{Evolution equation of the correlation function vector}
By making Fourier transform, Eq.~(\ref{LZ}) becomes
%%%%%%%%%%%%%%%%%%%%%%%%
\begin{equation}\label{LZk}
\L(\rho)=\sum_{k=-\pi}^{\pi} \Big( -i2J\cos k [\hat{n}_k,\rho]+(1-w)D^L_k(\rho)+(1+w)D^R_k(\rho) \Big),
\end{equation}
%%%%%%%%%%%%%%%%%%%%%%%%
where
%%%%%%%%%%%%%%%%%%%%%%%%
\begin{equation}
\begin{split}
& D^L_k(\rho)=2B_k\rho B^\dag_k-\{B^\dag_k B_k, \rho\}\\
&D^R_k(\rho)=2B^\dag_k\rho B_k-\{B_k B^\dag_k, \rho\},
\end{split}
\end{equation}
%%%%%%%%%%%%%%%%%%%%%%%%
and $B_k=\sqrt{\gamma_1}e^{ik}a^\dag_k+\sqrt{\gamma_2}a_{-k}$. Then, the expectation value of an arbitrary operator $\O$ meets the evolution equation:
%%%%%%%%%%%%%%%%%%%%%%%%%
\begin{equation}~\label{KO0}
\begin{split}
&\frac{d}{dt} \Tr (\O \rho(t)) = \sum_{k=-\pi}^{\pi} \Big( -i2J \cos k \Tr ([\O ,\hat{n}_k] \rho) \\
&+ (1-w) \big( \, \Tr ([B^\dag_k , \O ] B_k \rho  ) + \Tr (B^\dag_k [\O, B_k] \rho)  \, \big)  \Big.\\
&\Big. + (1+w) \big(\, \Tr ([B_k , \O ] B^\dag_k \rho  ) + \Tr (B_k [\O, B^\dag_k] \rho) \, \big) \Big).
\end{split}
\end{equation}
%%%%%%%%%%%%%%%%%%%%%%%%%

We define two-operator correlation functions in momentum space: $G_{k_1,\,k_2}=\Tr(a^\dag_{k_1} a_{k_2} \rho)$, $D_{k_1,\,k_2}=\Tr(a_{k_1} a_{k_2} \rho)$, and $D^*_{k_1,\,k_2}=\Tr(a^\dag_{k_2} a^\dag_{k_1} \rho)$. Substituting $\O = a^\dag_{k_1} a_{k_2}$, $\O = a^\dag_{-k_2} a_{-k_1}$, $\O = a_{k_2} a_{-k_1}$ and $\O = a^\dag_{-k_2} a^\dag_{k_1}$ into Eq.(\ref{KO0}), we find that the dynamical evolution is closed in terms of the correlation function vector
%%%%%%%%%%%%%%%%%%%%%%%%
\begin{equation}
\Psi_{k_1 k_2}=(G_{k_1,k_2},G_{-k_2,-k_1},D_{k_2,-k_1},D^*_{k_1,-k_2})^\T,
\end{equation}
%%%%%%%%%%%%%%%%%%%%%%%%
and the evolution equation is
%%%%%%%%%%%%%%%%%%%%%%%%
\begin{equation}~\label{dXk}
\frac{d}{dt} \Psi_{k_1 k_2} = \X_{k_1 k_2}\Psi_{k_1 k_2}+V_{k_1 k_2},
\end{equation}
%%%%%%%%%%%%%%%%%%%%%%%%
where
%%%%%%%%%%%%%%%%%%%%%%%%
\begin{eqnarray}~\label{XkVk}
 & \X_{k_1 k_2}  =-4\gamma\1\otimes \1 + i2J\cos k_1 \sigma_z \otimes \sigma_z -i2J\cos k_2 \1 \otimes \sigma_z \nonumber \\
              & +4\sqrt{\gamma_1 \gamma_2} \cos k_1 \sigma_x \otimes \sigma_z -4\sqrt{\gamma_1 \gamma_2} \cos k_2 \sigma_y \otimes \sigma_y
\end{eqnarray}
%%%%%%%%%%%%%%%%%%%%%%%%
and
%%%%%%%%%%%%%%%%%%%%%%%%
\begin{eqnarray}
\begin{split}
V_{k_1 k_2}  = \delta_{k_1,k_2} & \Big( 2\gamma + 2w(\gamma_2 - \gamma_1), \, 2\gamma + 2w(\gamma_2 -\gamma_1),\\
& i4w\sqrt{\gamma_1 \gamma_2 } \sin k_1,\  -i4w\sqrt{\gamma_1 \gamma_2 } \sin k_1 \Big)^{\T}
\end{split}
\end{eqnarray}
%%%%%%%%%%%%%%%%%%%%%%%%

The damping matrix $\X_{k_1k_2}$ has four eigenstates which fulfill the equation
%%%%%%%%%%%%%%%%%%%%%%%%
\begin{equation}
\X_{k_1k_2}|\Gamma_{k_1k_2}^{\pm \pm}\rangle=\Gamma_{k_1k_2}^{\pm \pm} |\Gamma_{k_1k_2}^{\pm \pm}\rangle,
\end{equation}
%%%%%%%%%%%%%%%%%%%%%%%%
with the eigenvalues given by
%%%%%%%%%%%%%%%%%%%%%%%%
\begin{equation}
\begin{split}
&\Gamma_{k_1k_2}^{\pm \pm}=-4\gamma \\
&\pm 2\sqrt{4\gamma_1 \gamma_2-J^2} \sqrt{(|\cos k_1|\pm |\cos k_2|\mathrm{sgn}(4\gamma_1 \gamma_2-J^2))^2},
\end{split}
\end{equation}
%%%%%%%%%%%%%%%%%%%%%%%%
where $\mathrm{sgn}(x)$ is a sign function.

\begin{figure}[htbp]\centering
\includegraphics[width=8.5cm]{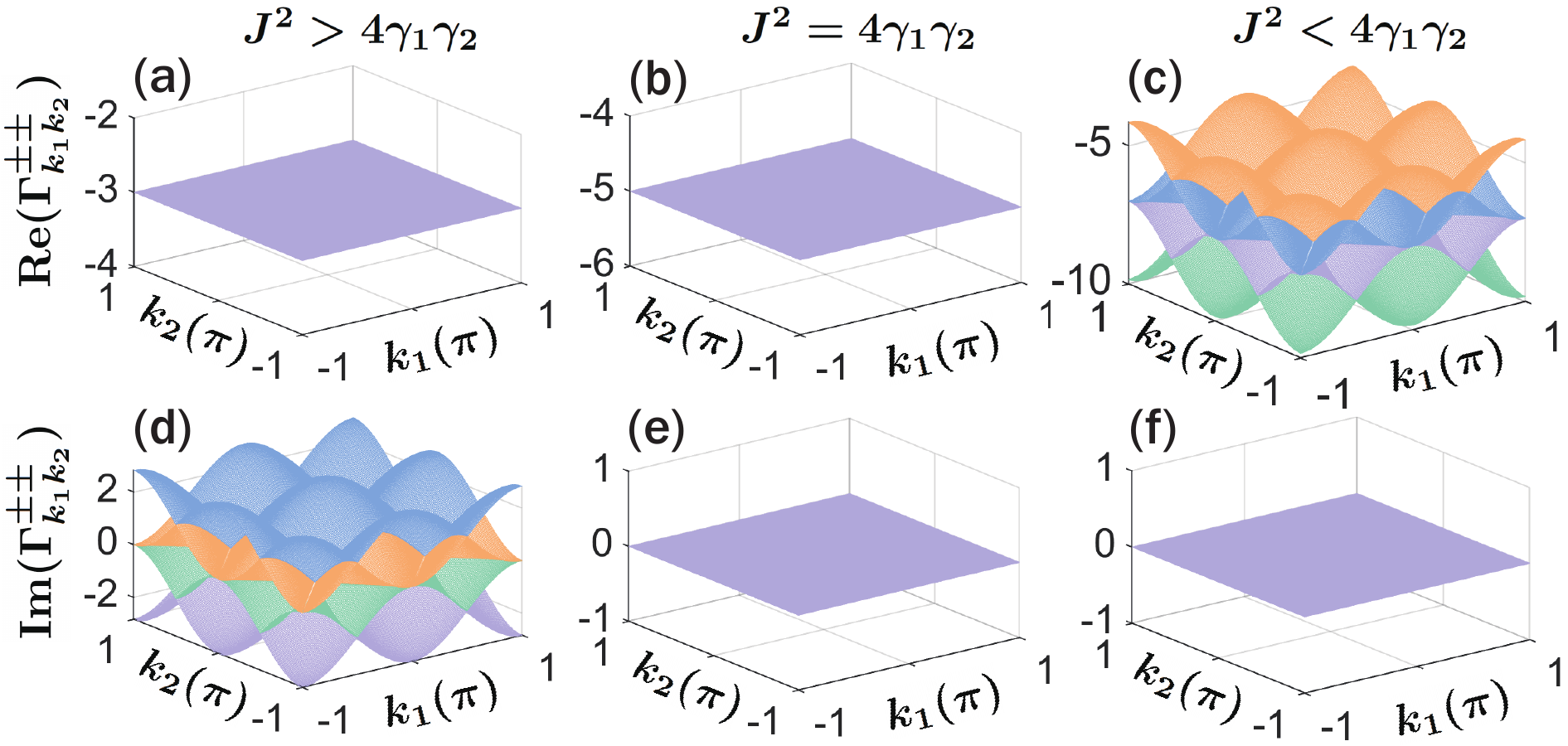}
\caption{(a)$\sim$(c) the real part of $\Gamma^{\pm \pm}_{k_1k_2}$. (d)$\sim$(f) the imaginary part of $\Gamma^{\pm \pm}_{k_1k_2}$. $J=1$ and $\gamma_1=0.25$ are for all subfigures. $\gamma_2=0.5$ in (a) and (d). $\gamma_2=1$ in (b) and (e). $\gamma_2=1.5$ in (c) and (f).}
\label{fig::4}
\end{figure}

Denoting $\k=(k_1, k_2)$, we show that the damping matrix $\X_{\k}$ has $\mathrm{TRS}$, $\mathrm{PHS}$, $\mathrm{CS}$, $\mathrm{IS}$ and $\mathrm{PTS}$:
%%%%%%%%%%%%%%%%%%%%%%%%
\begin{equation}~\label{SofX}
\begin{split}
&\mathrm{TRS}:\ U_{\mathcal{T}}\  \X^{*}_{\k}\  U_{\mathcal{T}}^{-1} = \X_{-\k}\ \Longrightarrow \ U_{\mathcal{T}}=\sigma_z \otimes \sigma_x; \ U_{\mathcal{T}} U_{\mathcal{T}}^{*}=1 \\
&\mathrm{PHS}:\ U_{\mathcal{C}}\  \X^{\T}_{\k}\  U_{\mathcal{C}}^{-1} = -\X_{-\k}\ \ \Longrightarrow \ \ U_{\mathcal{C}}=\1 \otimes \sigma_x ; \ U_{\mathcal{C}} U_{\mathcal{C}}^{*}=1 \\
&\mathrm{CS}:\ U_{\Gamma}\  \X^{\dag}_{\k}\  U_{\Gamma}^{-1} = -\X_{\k}\ \ \ \Longrightarrow \ \ \ U_{\Gamma} =\sigma_z \otimes \1; \ U_{\Gamma}^2=1 \\
&\mathrm{IS}:\ U_{\P}\  \X_{\k}\  U_{\P}^{-1} = \X_{-\k}\ \ \ \Longrightarrow \ \ \ U_{\P} =\1 \otimes \1 \\
&\mathrm{PTS}:\ U_{\PT}\  \X^{*}_{\k}\  U_{\PT}^{-1} = \X_{\k}\ \ \Longrightarrow \ \ U_{\PT}=\sigma_z \otimes \sigma_x.
\end{split}
\end{equation}
%%%%%%%%%%%%%%%%%%%%%%
Compared with the symmetry of the Liouvillian in Eq.(\ref{SofL}), $\X_{\k}$ has higher symmetry, which makes $\X_{k_1 k_2}$ have a similar band structure to $\hat{L}_k$. In Fig.~\ref{fig::4}, we see that $\Gamma_{k_1k_2}^{\pm \pm}$ fully inherits the dispersion characteristics of the real and imaginary part from the rapidity spectra in Fig.~\ref{fig::2}.
%%%%%%%%%%%%%%%%%%%%%%%%
%%%%%%%%%%%%%%%%%%%%%%%%

%%%%%%%%%%%%%%%%%%%%%%%%%%%%%%%%%%%%%%%%%%%%%%%%%%%%%%%%%%%%%%%%%%%%%%%%%%%%%%%%%%%%%%%%%%%%%%%%%%%%%%%%%%%%%%%%%%%%%%%%
%%%%%%%%%%%%%%%%%%%%%%%%%%%%%%%%%%%%%%%%%%%%%%%%%%%%%%%%%%%%%%%%%%%%%%%%%%%%%%%%%%%%%%%%%%%%%%%%%%%%%%%%%%%%%%%%%%%%%%%%

Setting Eq.~(\ref{dXk}) to $0$, we obtain the correlation functions of the steady state by
%%%%%%%%%%%%%%%%%%%%%%%%%
\begin{equation}
\left( G^s_{k_1,k_2} \ G^s_{-k_2,-k_1} \  D^s_{k_2, -k_1} \ D^{s*}_{k_1,-k_2 }  \right)^{\T} = - \X_{k_1 k_2}^{-1} V_{k_1 k_2},
\end{equation}
%%%%%%%%%%%%%%%%%%%%%%%%%
where the superscript $s$ represents the steady state expected value. When $k_1 = k_2=k$, we get the particle number distribution of the steady state in momentum space $n^s_k$:
%%%%%%%%%%%%%%%%%%%%%%%%%
\begin{equation}~\label{Gkk}
\begin{split}
&n^s_k=G^{s}_{kk}\\
&= \frac{(1-w)\gamma_1 + (1+w)\gamma_2 }{2\gamma} - \frac{2Jw\gamma_1 \gamma_2 \cos^2 k \sin k}{\gamma^3 + \gamma (J^2 - 4\gamma_1 \gamma_2) \cos^2 k}.
\end{split}
\end{equation}
%%%%%%%%%%%%%%%%%%%%%%%%%
Due to the translation invariance of our system, the particle number distributes uniformly on each site. Therefore, particle number on site $l$ in the thermodynamic limit can be calculated by
%%%%%%%%%%%%%%%%%%%%%%%%%
\begin{equation}
\begin{split}
&n^s_l= \frac{1}{L} \sum_{j=1}^{L} n^s_j =  \frac{1}{L} \sum_k n^s_k \\
&= \frac{1}{2\pi} \int_{k=-\pi}^{\pi} dk\  n^s_k = \frac{1}{2} + \frac{w(\gamma_2 - \gamma_1)}{2\gamma}.
\end{split}
\end{equation}
%%%%%%%%%%%%%%%%%%%%%%%%%

\begin{figure}[htbp]\centering
\includegraphics[width=8.5cm]{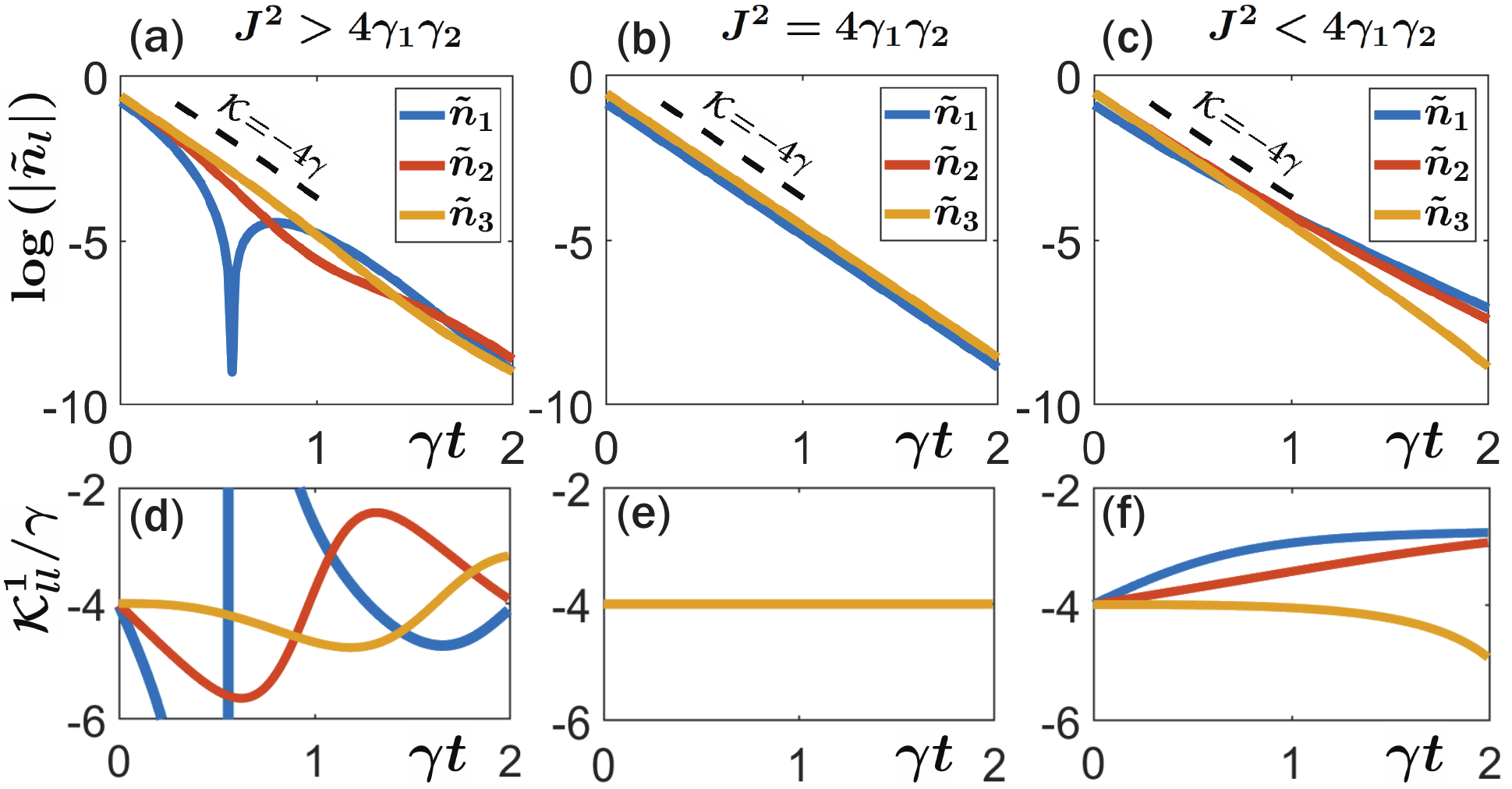}
\caption{The damping of particle number at different sites. The lattice has $15$ sites under the periodic boundary condition. Initial state is a single excitation on the first site from vacuum. The time evolutions of $\log(|\tilde{n}_l|)$ are shown in (a), (b), (c), and their derivatives $\mathcal{K}^{1}_{ll}$ are shown in (d), (e) and (f). The blue, red and orange lines are corresponding to $l=1$, $l=2$ and $l=3$, respectively. In (a) and (d), $\gamma_2$ is set as $0.5$. In (b) and (e), $\gamma_2=1$. In (c) and (f), $\gamma_2=1.5$. Others parameters are the same in all subfigures with $J=1$, $\gamma_1=0.25$ and $w=0.25$. The black dashed line represents a constant decay rate as $\tilde{n}_l \propto e^{-4\gamma t}$.}
\label{fig::5}
\end{figure}

%%%%%%%%%%%%%%%%%%%%%%%%%%%%%%%%%%%%%%%%%%%%%%%%%%%%%%%%%%%%%%%%%%%%%%%%%%%%%%%%%%%%%%%%%%%%%%%%%%%%%%%%%%%%%%%%%%%%%%%%
%%%%%%%%%%%%%%%%%%%%%%%%%%%%%%%%%%%%%%%%%%%%%%%%%%%%%%%%%%%%%%%%%%%%%%%%%%%%%%%%%%%%%%%%%%%%%%%%%%%%%%%%%%%%%%%%%%%%%%%%
%\section{Flat-band damping dynamics}
Damping dynamics displays the converging processes from initial state to NESS~\cite{Wangzhong2019}. Here, we show that the ``flat band'' in the real or imaginary part or both parts will effectively influence the damping behaviors in real space.
We concentrate on the vector
%%%%%%%%%%%%%%%%%%%%%%%%
\begin{equation}
\Psi_{l_1 l_2}=(G_{l_1,l_2}, G_{l_2,l_1}, D_{l_2,l_1}, D^*_{l_1,l_2})^{\T},
\end{equation}
%%%%%%%%%%%%%%%%%%%%%%%%
consisting of real-space correlation functions:
%%%%%%%%%%%%%%%%%%%%%%%%
$$
G_{l_1,\,l_2}=\Tr(a^\dag_{l_1} a_{l_2} \rho),\ \ D_{l_1,\,l_2}=\Tr(a_{l_1} a_{l_2} \rho),\ \ D^*_{l_1,\,l_2}=\Tr(a^\dag_{l_2} a^\dag_{l_1} \rho).
$$
%%%%%%%%%%%%%%%%%%%%%%%%
Introduce the deviating expectation of operator $\hat{O}$ as
%%%%%%%%%%%%%%%%%%%%%%%%
\begin{equation}
\widetilde{O}(t)=\langle \hat{O} \rangle (t)-\langle \hat{O} \rangle^S,
\end{equation}
%%%%%%%%%%%%%%%%%%%%%%%%
to describe the deviation from the steady state expectation value $\langle \hat{O} \rangle^S=\langle \hat{O} \rangle (\infty)$. From Eq.~(\ref{dXk}), we get
%%%%%%%%%%%%%%%%%%%%%%%%
\begin{equation}
\frac{d}{dt} \widetilde{\Psi}_{k_1 k_2} = \X_{k_1 k_2} \widetilde{\Psi}_{k_1 k_2},
\end{equation}
%%%%%%%%%%%%%%%%%%%%%%%%
where $\widetilde{\Psi}_{k_1 k_2}(t)=\Psi_{k_1 k_2}(t)-\Psi_{k_1 k_2}(t=\infty)$. Making Fourier transformation, we have
%%%%%%%%%%%%%%%%%%%%%%%%
\begin{equation}
\widetilde{\Psi}_{l_1 l_2}(t)=\sum_{k_1 k_2} e^{i(-k_1l_1 +k_2 l_2)}\widetilde{\Psi}_{k_1 k_2}(t).
\end{equation}
%%%%%%%%%%%%%%%%%%%%%%%%
Decomposing an arbitrary initial state $\widetilde{\Psi}_{k_1 k_2}(0)$ by the eigenstates of $\X_{k_1 k_2}$ i.e.
%%%%%%%%%%%%%%%%%%%%%%%%
\begin{equation}
\widetilde{\Psi}_{k_1 k_2}(0)=\sum_{\alpha \beta} C^{\alpha \beta}_{k_1 k_2} |\Gamma^{\alpha \beta}_{k_1 k_2}\rangle,
\end{equation}
%%%%%%%%%%%%%%%%%%%%%%%%
where $\alpha$ and $\beta$ take $\pm$, we have
%%%%%%%%%%%%%%%%%%%%%%%%
\begin{equation}\label{PSIt}
\widetilde{\Psi}_{l_1 l_2}(t)=\sum_{\k,\u} e^{i\k \cdot \r} C^{\u}_{\k} e^{t\,\Gamma^{\u}_{\k}} |\Gamma^{\u}_{\k}\rangle,
\end{equation}
%%%%%%%%%%%%%%%%%%%%%%%%
where $\k=(k_1,\, k_2)$, $\r=(-l_1,\, l_2)$ and $\u=(\alpha,\,\beta)$. For non-zero Liouvillian gap, the system exponentially decays to NESS in a long-time limit, i.e. $\widetilde{\rho} \propto e^{-\kappa t}$. We can define instantaneous decay rate $\K (t)$ for all-time behavior, i.e. $\widetilde{\rho} \propto e^{-\K (t) t}$, in which ${\rm lim}_{t \rightarrow \infty} \K (t) = \kappa$. The instantaneous decay rate of the $j$ component of $\widetilde{\Psi}_{l_1 l_2}(t)$ is defined as
%%%%%%%%%%%%%%%%%%%%%%%%
\begin{equation}
\K^j_{l_1 l_2}=\frac{d}{dt}\log \left(| \widetilde{\Psi}^j_{l_1 l_2}(t) |\right).
\end{equation}
%%%%%%%%%%%%%%%%%%%%%%%%
Below we unveil how $\K (t)$ is affected by the dispersion of $\Gamma^{\u}_{\k}$ through Fig.~\ref{fig::5}, in which the damping behaviors of local deviating particle number $\widetilde{n}_l=\widetilde{G}_{ll}$ from the initial state with a single excitation on site $1$ are shown:

(i) When FB appears, $\Gamma^{\u}_{\k}$ becomes a constant, denoted by $\Gamma_0$. Then we have
%%%%%%%%%%%%%%%%%%%%%%%%
\begin{eqnarray}
&\widetilde{\Psi}_{l_1 l_2}(t)=e^{\Gamma_0 t} \sum_{\k,\u} e^{i\k \cdot \r} C^{\u}_{\k} |\Gamma^{\u}_{\k}\rangle,\\
&\K^j_{l_1 l_2}(t)=\Re(\Gamma_0),
\end{eqnarray}
%%%%%%%%%%%%%%%%%%%%%%%%
which means for arbitrary initial state different two-operator correlation functions will synchronously relax to their steady state expectation values with the same decay rate, as demonstrated in Fig.~\ref{fig::5} (b) and (e), where different curves of $\log(\tilde{n}_l)$ as a function with $\gamma t$ have the same constant slope, i.e. $\K^1_{ll}=4\gamma$ for all $l$.

(ii) When $\Gamma^{\u}_{\k}$ is only dispersionless in its real part, we set $\Gamma^{\u}_{\k}=-x_0-iy^{\u}(\k)$, where $x_0$ and $y^{\u}(\k)$ are real. Then we have
%%%%%%%%%%%%%%%%%%%%%%%%
\begin{eqnarray}
&\widetilde{\Psi}_{l_1 l_2}(t)=e^{-x_0 t}\sum_{\k,\u} C^{\u}_{\k} e^{i\k \cdot \r} e^{-iy^{\u}(k)t} |\Gamma^{\u}_{\k}\rangle,\\
&\K^j_{l_1 l_2}=-x_0 + \frac{d}{dt} \log \left( \big|\sum_{\k,\u} C^{\u}_{\k}|\Gamma^{\u}_{\k}\rangle_j e^{i\big(\k \cdot \r -y^{\u}(\k)t \big)} \big| \right).\label{KI}
\end{eqnarray}
%%%%%%%%%%%%%%%%%%%%%%%%
The right side of Eq.~(\ref{KI}) contains sum of a series of plane waves, which leads to $\K^j_{l_1 l_2} (t)$ oscillating around $x_0$, as shown in Fig.~\ref{fig::5} (d). The oscillating slopes lead to continuously intersecting curves in Fig.~\ref{fig::5} (a).

(iii) When $\Gamma^{\u}_{\k}$ is only dispersionless in its imaginary part, we set $\Gamma^{\u}_{\k}=-(x_c + \delta x^{\u}(\k))-iy_0$, where $x_c$ and $\delta x^{\u}(\k)$ are the central value and the offset function of $\Re (\Gamma^{\u}_{\k})$, and $y_0$ is the imaginary part. Then we have
%%%%%%%%%%%%%%%%%%%%%%%%
\begin{eqnarray}
&\widetilde{\Psi}_{l_1 l_2}(t)=e^{-(x_c + iy_0)t}\sum_{\k,\u} C^{\u}_{\k} e^{i\k \cdot \r} e^{-\delta x^{\u}(k)t} |\Gamma^{\u}_{\k}\rangle,\\
&\K^j_{l_1 l_2}=-x_c + \frac{d}{dt} \log \left( \big|\sum_{\k,\u} C^{\u}_{\k}|\Gamma^{\u}_{\k}\rangle_j e^{i\k \cdot \r} e^{-\delta x^{\u}(\k)t}\big| \right).\label{KR}
\end{eqnarray}
%%%%%%%%%%%%%%%%%%%%%%%%
Since $\delta x^{\u}(\k)$ is real, the relaxation process does not display oscillating decay rates (see Fig.~\ref{fig::5} (f)). This induces the forked damping curves typically as shown in Fig.~\ref{fig::5} (c).
%%%%%%%%%%%%%%%%%%%%%%%%
%%%%%%%%%%%%%%%%%%%%%%%%

The above damping dynamics is directly related to dispersion of damping-matrix spectra. The damping-matrix spectra reflect the decay of correlation functions, however, the Liovillian spectra reflect the decay of the whole system. We prove that for real physical processes the damping-matrix spectra are included in Liouvillian spectra in Appendix \ref{AP:5}. Therefore, for more general models with closed evolution equations of two-operator correlation functions, the dispersionless Liouvillian bands will lead to dispersionless damping-matrix spectra, and then give rise to the same dynamical signatures as shown in our model.

%%%%%%%%%%%%%%%%%%%%%%%%%%%%%%%%%%%%%%%%%%%%%%%%%%%%%%%%%%%%%%%%%%%%%%%%%%%%%%%%%%%%%%%%%%%%%%%%%%%%%%%%%%%%%%%%%%%%%%%%
%%%%%%%%%%%%%%%%%%%%%%%%%%%%%%%%%%%%%%%%%%%%%%%%%%%%%%%%%%%%%%%%%%%%%%%%%%%%%%%%%%%%%%%%%%%%%%%%%%%%%%%%%%%%%%%%%%%%%%%%
\section{Compact localized normal master modes and dynamic localization}
In isolated system, FBs lead to localized eigenstates by destructive interference. Now, we exactly solve our model (see Appendix \ref{AP:3} for details) and show that the LFB can induce dynamic localization by compact localized normal master modes (CLNMMs), which suppress propagation of local perturbation on NESS.

Usually, the odd-parity part of $\hat{L}$ has no effect on the expectation value of observation in a pure fermionic system (see Appendix \ref{AP:3:3}). Therefore, we focus on the balanced model ($w=0$) with even parity ($P=1$), whose Liouvillian is illustrated in Fig.~\ref{fig::1} (d). By solving the equation $\zeta_i(k) |\Omega\rangle=0$ for $i=1 \sim 4$, we get the steady state $|\Omega\rangle$ as
%%%%%%%%%%%%%%%%%%%%%%%%%
\begin{equation}~\label{SS}
|\Omega\rangle=\frac{1}{\N}\prod_{k=-\pi}^{\pi} (1+a^\dag_k c^\dag_{-k}) |0\rangle=\frac{1}{\N}\prod_{l=1}^{L} (1+a^\dag_l c^\dag_l) |0\rangle,
\end{equation}
%%%%%%%%%%%%%%%%%%%%%%%%%
where $\N=2^L$ and this state is independent with $\gamma_1$ and $\gamma_2$. At the FB point with $J=2\sqrt{\gamma_1 \gamma_2}$, the exceptional degeneracy occurs in the non-Hermitian matrix $\L_k$ of Eq.~(\ref{Lk1}) with four eigenstates coalescing into two. Then $\hat{L}_k$ is reduced to
%%%%%%%%%%%%%%%%%%%%%%%%
\begin{equation}
\hat{L}_k=-2\gamma \left(\zeta^{'}_A(k) \zeta_A(k)+\zeta^{'}_B(k) \zeta_B(k)\right),
\end{equation}
%%%%%%%%%%%%%%%%%%%%%%%%
where
%%%%%%%%%%%%%%%%%%%%%%%%%
\begin{equation}%
\begin{split}
& \zeta^{'}_A(k)=-a^\dag_k + c_{-k},\ \ \ \zeta_A(k) =\frac{1}{2}(-a_k+ic_k+ia^\dag_{-k}+c^\dag_{-k}),\\
& \zeta^{'}_B(k)= a_k + c^\dag_{-k},\ \ \ \zeta_B(k)=\frac{1}{2}(a^\dag_k - ic^\dag_k + i a_{-k} +c_{-k}).
\end{split}
\end{equation}
%%%%%%%%%%%%%%%%%%%%%%%%%
Making Fourier transformation, we get
%%%%%%%%%%%%%%%%%%%%%%%%
\begin{equation}
\hat{L}=\sum_l -2\gamma [\zeta^{'}_A(l) \zeta_A(l) + \zeta^{'}_B(l) \zeta_B(l) ],
\end{equation}
%%%%%%%%%%%%%%%%%%%%%%%%
where
%%%%%%%%%%%%%%%%%%%%%%%%%
\begin{equation}
\begin{split}
&\zeta^{'}_A(l)=\sum_k e^{-ikl}\zeta^{'}_A(k)=-a^\dag_l + c_l, \\
&\zeta^{'}_B(l)=\sum_k e^{ikl}\zeta^{'}_B(k)=a_l + c^\dag_l.
\end{split}
\end{equation}
%%%%%%%%%%%%%%%%%%%%%%%%%
We coin $\zeta^{'}_{A,B}(l) |\Omega\rangle$ as CLNMM states since they are eigen-modes of $\hat{L}$  and $\zeta^{'}_{A,B}(l)$ changes NESS locally.

We can also understand CLNMMs intuitively from the perspective of destructive interference. Writing the real-space Liouvillian with $w=0$, $J=2\sqrt{\gamma_1 \gamma_2}=1$ as
%%%%%%%%%%%%%%%%%%%%%%%%
\begin{equation}
\hat{L}=\sum_l (\hat{h}_l + \hat{f}_l-2\gamma),
\end{equation}
%%%%%%%%%%%%%%%%%%%%%%%%
where the hopping term $h_l$ is defined as
%%%%%%%%%%%%%%%%%%%%%%%%
\begin{equation}
\hat{h}_l=-i(a^\dag_{l+1}a_l+h.c.)+i(c^\dag_{l+1}c_l+h.c.)-(a^\dag_{l+1}c_l + c^\dag_{l+1}a_l + h.c),
\end{equation}
%%%%%%%%%%%%%%%%%%%%%%%%
and the pairing term $\hat{f}_l$ is defined as
%%%%%%%%%%%%%%%%%%%%%%%%
\begin{equation}
f_l=2\gamma \,(a^\dag_l c^\dag_l +c_l a_l),
\end{equation}
%%%%%%%%%%%%%%%%%%%%%%%%
we can check that
%%%%%%%%%%%%%%%%%%%%%%%%
\begin{equation}
\hat{f}_l\, a_l\,  |\Omega\rangle=\hat{f}_l\, c_l\,  |\Omega\rangle=\hat{f}_l\, a^\dag_l\,  |\Omega\rangle=\hat{f}_l\, c_l^\dag\,  |\Omega\rangle=0.
\end{equation}
%%%%%%%%%%%%%%%%%%%%%%%%
This implies that the pairing terms do not affect a single particle or hole excited on the NESS. Therefore, for these states only hopping terms make sense. We schematically plot this reduced ladder in Fig.~\ref{fig::1} (e). It is easy to find another created operator of CLNMM as
%%%%%%%%%%%%%%%%%%%%%%%%
\begin{equation}
\zeta^{'}_C(l)=a^\dag_l - ic^\dag_l,
\end{equation}
%%%%%%%%%%%%%%%%%%%%%%%%
from the view of destructive interference, which forbids the state $\zeta^{'}_C(l)|\Omega$  transferring to other sites. We can also check that
%%%%%%%%%%%%%%%%%%%%%%%%
\begin{equation}
\hat{L}\ \zeta^{'}_C(l)\, |\Omega\rangle=-2\gamma\ \zeta^{'}_C(l)\, |\Omega\rangle.
\end{equation}
%%%%%%%%%%%%%%%%%%%%%%%%
%%%%%%%%%%%%%%%%%%%%%%%%
\begin{figure}[htbp]\centering
\includegraphics[width=8.5cm]{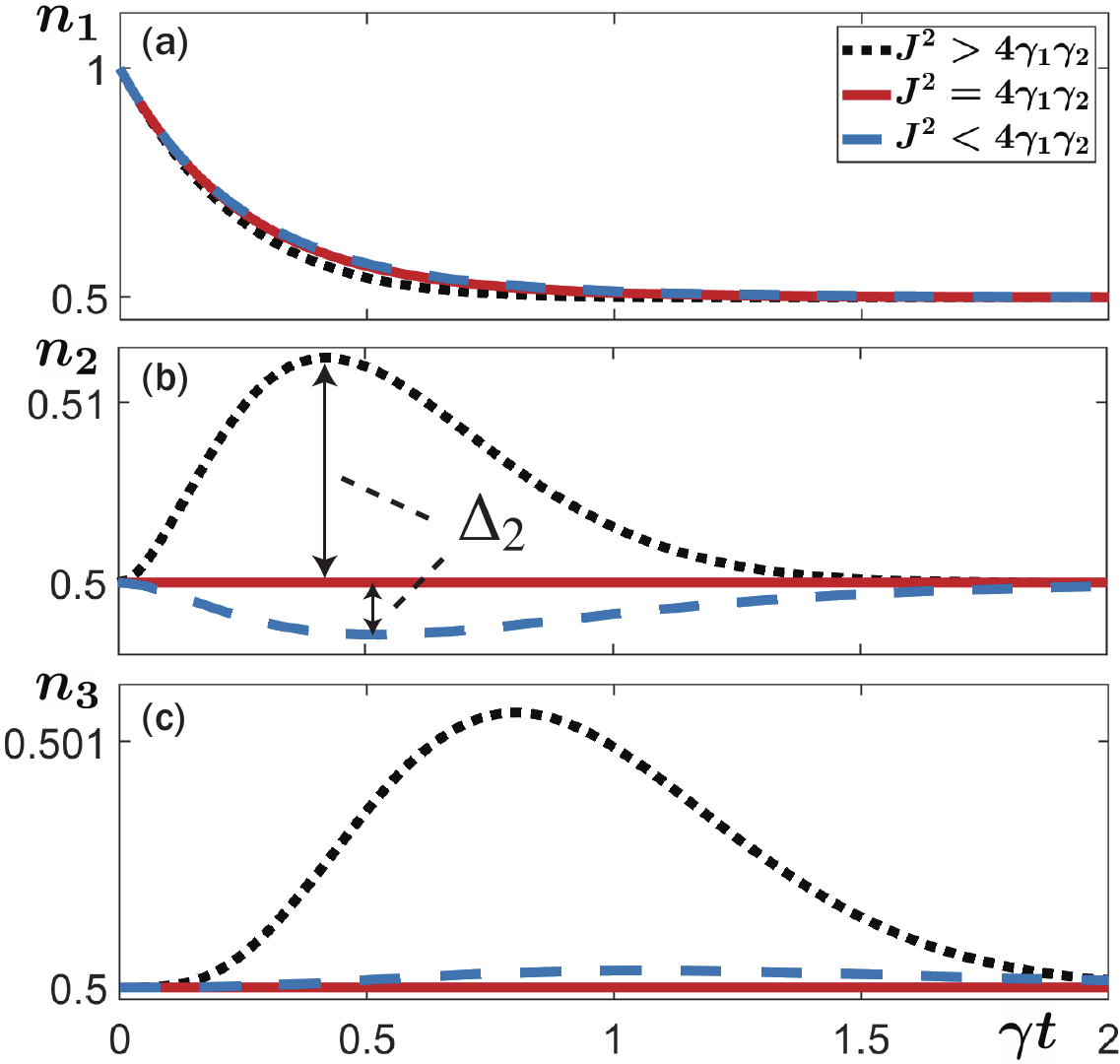}
\caption{The time evolution of particle number on the first site $n_1$ shown in (a), second site $n_2$ in (b) and third site $n_3$ in (c). Initial state is $a_1^\dag \rho_{s} a_1 / \Tr (a_1^\dag \rho_{s} a_1)$ corresponding to a quantum jump on the first site of steady state. The periodic lattice has $15$ sites with $w=0$, $J=1$, $\gamma_1=0.25$ in all subfigures. The black dotted, red solid, and blue dashed lines are corresponding to $\gamma_2=0.5$, $\gamma_2=1$ and $\gamma_2=1.5$, respectively. $\Delta_2$ (Eq. (56)) is the maximum deviation of the particle number on the second site from NESS.}
\label{fig::6}
\end{figure}
%%%%%%%%%%%%%%%%%%%%%%%%

The CLNMMs contain decay information of quantum jumps. To see it clearly, we map the $\CC -$representation state $\zeta^{'}_A (l)\zeta^{'}_B (l)|\Omega\rangle$, for example, back to  density-matrix representation:
%%%%%%%%%%%%%%%%%%%%%%%%%
\begin{equation}\label{ZAZB}
\zeta^{'}_A (l)\zeta^{'}_B (l)|\Omega\rangle \to -a^\dag_l a_l \rho_{s} + \rho_{s} a_l a^\dag_l + a_l \rho_{s} a^\dag_l - a^\dag_l \rho_{s} a_l,
\end{equation}
%%%%%%%%%%%%%%%%%%%%%%%%%
where $\rho_s$ is the density matrix of NESS. The terms $a_l \rho_{s} a^\dag_l$ and $a^\dag_l \rho_{s} a_l$ are exactly corresponding to local quantum jumps on NESS. Since the decay mode in Eq.~(\ref{ZAZB}) includes only  operators on the local site, it implies that the local perturbation on NESS will locally decay to NESS. To show it clearly, we simulate the time evolution from an initial state
%%%%%%%%%%%%%%%%%%%%%%%%
\begin{equation}
\rho_0=\frac{a_1^\dag \rho_{s} a_1 }{\Tr (a_1^\dag \rho_{s} a_1)},
\end{equation}
%%%%%%%%%%%%%%%%%%%%%%%%
in Fig.~\ref{fig::6}, where $\rho_0$ represents a quantum jump on the first site of NESS. In the beginning, the jump makes the first site particle number $n_1$ increase to 1 and the particle number of others sites keep their steady state value $0.5$. The red solid line, black dotted line and blue dashed line are corresponding to the situation with $J^2=4\gamma_1 \gamma_2$ (LFB), $J^2>4\gamma_1 \gamma_2$, and $J^2<4\gamma_1 \gamma_2$, respectively. We can see that when $J^2 \ne 4\gamma_1 \gamma_2$, the perturbation can spread from $n_1$ to $n_3$.  However, for the case with LFB, the perturbation excitation decays locally without going through to $n_2$ and $n_3$, indicating the occurrence of dynamical localization.

%%%%%%%%%%%%%%%%%%%%%%%%
\begin{figure}[htbp]\centering
\includegraphics[width=8.5cm]{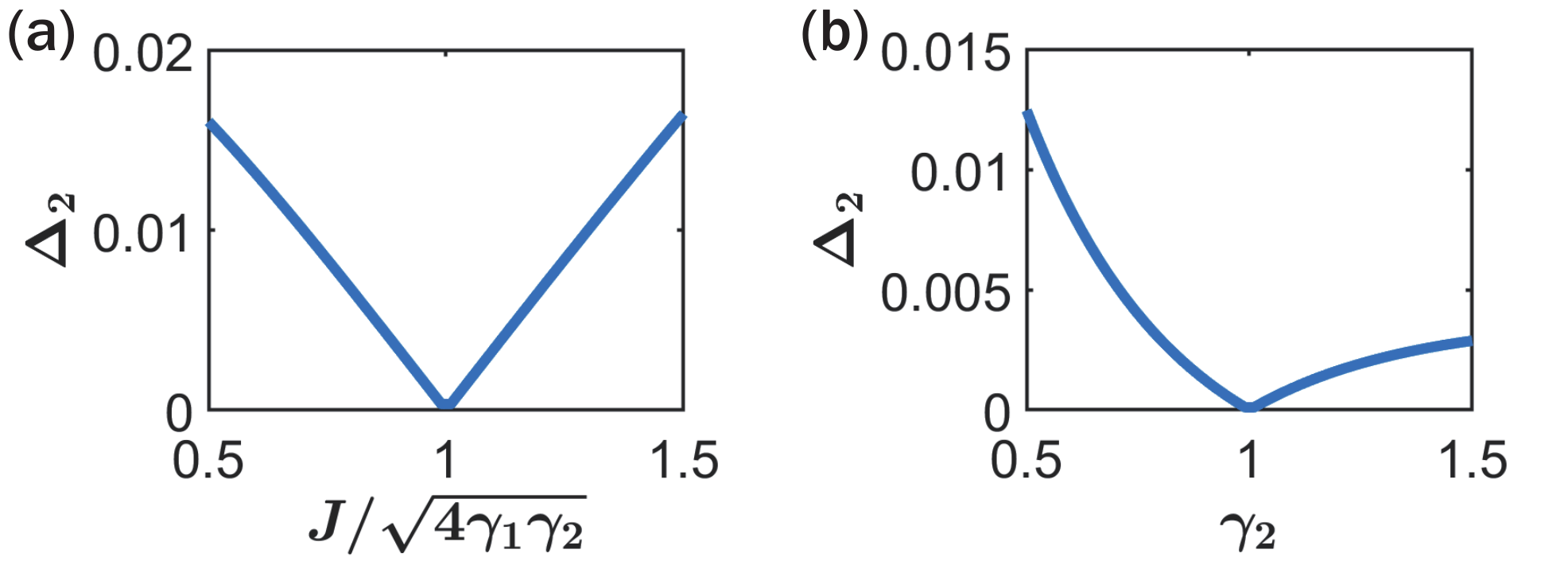}
\caption{ The maximum deviation of the second site particle number $\Delta_2$ as a function of $J$ in (a) and $\gamma_2$ in (b). Both curves are calculated in a 15-site periodic lattice with $w=0$. In (a), $\gamma_1=\gamma_2=0.5$. In (b), $J=1$ and $\gamma_1=0.25$. In both curves, $\Delta_2$ reaches $0$ at the flat-band condition $J^2=4 \gamma_1 \gamma_2$.}
\label{Fig::7}
\end{figure}
%%%%%%%%%%%%%%%%%%%%%%%%
Furthermore, we use the maximum deviation of the second site particle number
%%%%%%%%%%%%%%%%%%%%%%%%
\begin{equation}
\Delta_2={\rm max}(|\widetilde{n}_2(t)|)={\rm max}(|n_2(t)-n_2(\infty)|),
\end{equation}
%%%%%%%%%%%%%%%%%%%%%%%%
to reflect the ability of propagating perturbation. Small $\Delta_2$ is corresponding to weak propagation ability of perturbation. Values of $\Delta_2$ for different parameters are shown in Fig.~\ref{Fig::7}, where only in the LFB case, $\Delta_2$ equals $0$. It reflects the LFB can suppress the propagation of perturbation.

\section{Summary}
We construct flat-band models in open system with correlated gain and loss and demonstrate that the Liouvillian dispersion can affect the damping dynamics of the local particle number, intermediated by the damping matrix of the correlation function vector. When the Liouvillian flat band appears, the particle number in different sites will relax to their stable values synchronously. When only the real or imaginary part of the rapidity spectrum is dispersionless, the damping behaviors show oscillating or forked characteristics. Furthermore, we show that a Liouvillian flat band can induce dynamical localization on NESS by the compact localized normal master modes.

\section*{Acknowledgements}
We thank X. L. Wang, Z. Y. Zheng and C. X. Guo for helpful discussions.The work is supported by National Key Research and Development Program of China (Grant No.2021YFA1402104), the NSFC under Grants No. 12174436 and No. T2121001, and the Strategic Priority Research Program of Chinese Academy of Sciences under Grant No.XDB33000000.

%%%%%%%%%%%%%%%%%%%%%%%%%%%%%%%%%%%%%%%%%%%%%%%%%%%%%%%%%%%%%%%%%%%%%%%%%%%%%%%%%%%%%%%%%%%%%%%%%%%%%%%%%%
%%%%%%%%%%%%%%%%%%%%%%%%%%%%%%%%%%%%%%%%%%%%%%%%%%%%%%%%%%%%%%%%%%%%%%%%%%%%%%%%%%%%%%%%%%%%%%%%%%%%%%%%%%
\appendix
%%%%%%%%%%%%%%%%%%%%%%%%%%%%%%%%%%%%%%%%%%%%%%%%%%%%%%%%%%%%%%%%%%%%%%%%%%%%%%%%%%%%%%%%%%%%%%%%%%%%%%%%%%
%%%%%%%%%%%%%%%%%%%%%%%%%%%%%%%%%%%%%%%%%%%%%%%%%%%%%%%%%%%%%%%%%%%%%%%%%%%%%%%%%%%%%%%%%%%%%%%%%%%%%%%%%%
\section{Mapping of Lindblad master equation} \label{AP:1}
%%%%%%%%%%%%%%%%%%%%%%%%
\begin{figure*}[htbp]\centering
\includegraphics[width=15cm]{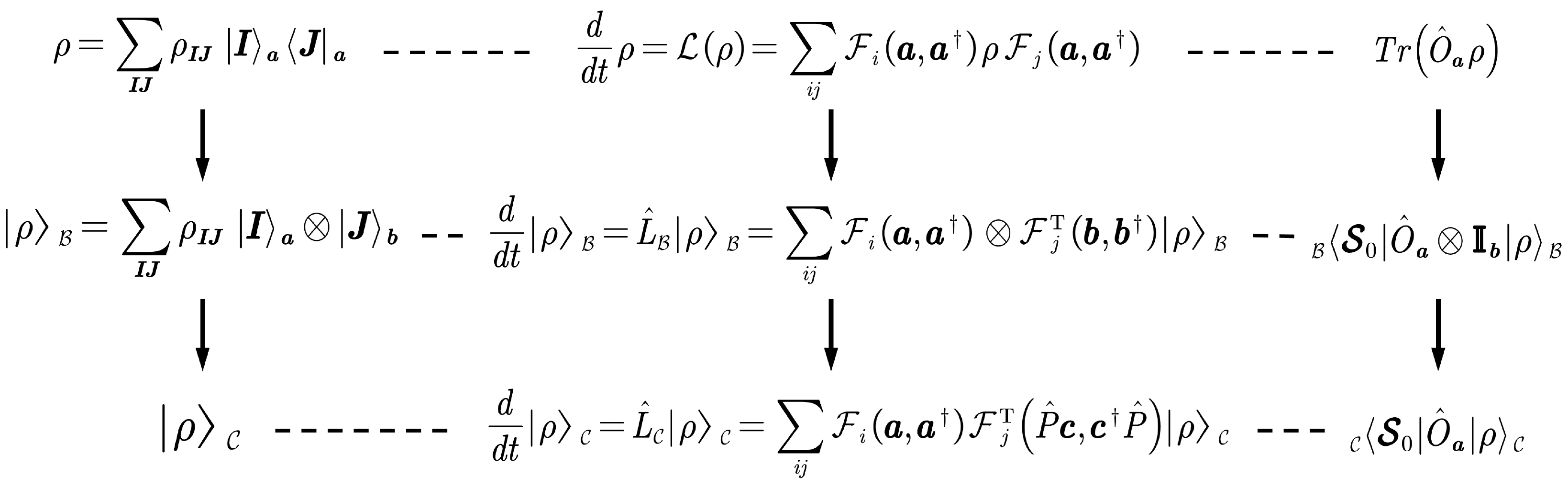}
\caption{Mapping of Lindblad master equation.}
\label{fig::8}
\end{figure*}
%%%%%%%%%%%%%%%%%%%%%%%%
The Lindblad master equation, the formalized density matrix $\rho$ and the Liouvillian superoperator $\L$ are shown in Eqs.~(\ref{Liouvillian}) and (\ref{formal}). First we carry out the Choi-Jamiolkowski isomorphism~\cite{Choi1975,Jamiolkowski1972,Tyson2003,Zwolak2004} to map the fermionic LME into representation $\B$ as
%%%%%%%%%%%%%%%%%%%%%%%%
\begin{equation}
\frac{d}{dt} |\,\rho\rangle_{\B} = \hat{L}_{\B} |\,\rho\rangle_{\B},
\end{equation}
%%%%%%%%%%%%%%%%%%%%%%%%
where $|\,\rho\rangle_{\B}$ is vectorized from $\rho$ and $\hat{L}_{\B}$ is mapped from $\L$. Specifically, the mapping is
%%%%%%%%%%%%%%%%%%%%%%%%
\begin{subequations}\label{M1}
\begin{align}
& \rho \to |\,\rho\rangle_{\B}=\sum_{\I \J} \rho_{\I \J}|\I\rangle_{\a} \otimes  |\J\rangle_{\b},\\
& \L \to \hat{L}_{\mathcal{B}}=\sum_{ij}\F_i(\a,\a^\dag)\otimes \F_j^\T (\b,\b^\dag),
\end{align}
\end{subequations}
%%%%%%%%%%%%%%%%%%%%%%%%
where $\b=(b_1, b_2, \cdots)$ is the set of annihilation operators of $b-$fermions, which is one-to-one mapping from $\a$, and $\T$ denotes matrix transpose. $|\I\rangle_{\a}$ and $|\J\rangle_{\b}$ are defined as
%%%%%%%%%%%%%%%%%%%%%%%%
\begin{subequations}
\begin{align}
&|\I\rangle_{\a} = (a^\dag_1)^{I_1}(a^\dag_2)^{I_2}\cdots (a^\dag_L)^{I_L} |0\rangle_{\a},\\
&|\J\rangle_{\b}=(b^\dag_1)^{J_1}(b^\dag_2)^{J_2}\cdots (b^\dag_L)^{J_L} |0\rangle_{\b},
\end{align}
\end{subequations}
%%%%%%%%%%%%%%%%%%%%%%%%
where $|0\rangle_{\a}$ and $|0\rangle_{\b}$ are vacuum state of all $a-$fermions and $b-$fermions, respectively. In this representation, the expectation value of the observable becomes
%%%%%%%%%%%%%%%%%%%%%%%%
\begin{equation}~\label{EB}
\langle \O_{\a} \rangle = _{\B}\langle \SS | \O_{\a} \otimes \1_{\b} |\,\rho\rangle_{\B},
\end{equation}
%%%%%%%%%%%%%%%%%%%%%%%%
where $_{\B} \langle \SS |$ is a special state defined as:
%%%%%%%%%%%%%%%%%%%%%%%%
\begin{equation}
\begin{split}
& _{\B}\langle \SS |= \sum_{\S} \langle \S |_{\a} \otimes \langle \S |_{\b} \\
& = \sum_{\S} \Big( \langle 0 |_{\a} (a_L)^{S_L}\cdot \cdot (a_1)^{S_1} \otimes \langle 0 |_{\b} (b_L)^{S_L}\cdot \cdot (b_1)^{S_1} \Big),
\end{split}
\end{equation}
%%%%%%%%%%%%%%%%%%%%%%%%
and $\1_{\b}$ is a unit operator of all $b-$fermions. The element $S_i$ of $\S=(S_1,S_2,\cdots)$ can take $0$ or $1$, and $\sum_{\S}$ requires a sum over all possible configurations of $\S$.
Let us prove Eq.~(\ref{EB}):
%%%%%%%%%%%%%%%%%%%%%%%%
\begin{equation}
\begin{split}
& \langle \O_{\a} \rangle = \sum_{\I \J \S} \rho_{\I \J}\  \langle \S |_{\a} \O_{\a} | \I \rangle_{\a} \langle \S |_{\b} \1_{\b} | \J \rangle_{\b} \\
                        &= \sum_{\I \J \S}    \rho_{\I \J}\  {_{\a}}\!\langle 0 | a_L^{S_L} \cdots a_1^{S_1} \O_{\a} (a^\dag_1)^{I_1} \cdots  (a^\dag_L)^{I_L} |0\rangle_{\a}\  \delta_{\S \J}   \\
                        &= \sum_{\I \J} \rho_{\I \J} \  {_{\a}}\!\langle 0 | a_L^{J_L} \cdots a_1^{J_1} \O_a (a^\dag_1)^{I_1} \cdots  (a^\dag_L)^{I_L} |0\rangle_{\a}\\
                        &= \sum_{\I \J} \rho_{\I \J} \  {_{\a}}\!\langle \J | \O_{\a}  | \I \rangle_{\a} =\Tr(\O_{\a} \rho).
\end{split}
\end{equation}
%%%%%%%%%%%%%%%%%%%%%%%%
In representation $\B$, operators satisfy the following relations:
%%%%%%%%%%%%%%%%%%%%%%%%
\begin{equation}~\label{CR}
\begin{split}
& \{a_i,a^\dag_j\}=\{b_i,b^\dag_j\}=\delta_{ij},\\
& \{a^\dag_i,a^\dag_j\}=\{a_i,a_j\}=\{b^\dag_i,b^\dag_j\}=\{b_i,b_j\}=0,\\
& [a^\dag_i,b_j]=[a_i^\dag,b^\dag_j]=[a_i,b_j]=[a_i,b^\dag_j]=0.
\end{split}
\end{equation}
%%%%%%%%%%%%%%%%%%%%%%%%
The commutation relations in Eq.~(\ref{CR}) are from the direct product between $a-$fermions and $b-$fermions, which are unfavorable for further analysis. To enforce fermionic anticommutation relations over all operators, we define operators of $c-$fermions as $\c^\dag=\b^\dag \hat{P}$ and $\c=\hat{P}\b$, where $\hat{P}$ is a parity operator defined as
%%%%%%%%%%%%%%%%%%%%%%%%
\begin{equation}
\hat{P}:=\exp \Big( i\pi\sum_l(a^\dag_l a_l + b^\dag_l b_l) \Big)=\exp \Big( i\pi\sum_l(a^\dag_l a_l + c^\dag_l c_l) \Big).
\end{equation}
%%%%%%%%%%%%%%%%%%%%%%%
It is easy to check the fermionic anticommutation relations in $a-$fermions and $c-$fermions:
%%%%%%%%%%%%%%%%%%%%%%%%
\begin{equation}
\begin{split}
& \{c_i,c^\dag_j\}=\delta_{ij},\ \ \ \ \{c^\dag_i,c^\dag_j\}=\{c_i,c_j\}=0, \\
& \{a^\dag_i,c_j\}=\{a_i^\dag,c^\dag_j\}=\{a_i,c_j\}=\{a_i,c^\dag_j\}=0
\end{split}
\end{equation}
%%%%%%%%%%%%%%%%%%%%%%%
By $\c$ we can fully fermionize system from representation $\B$ to representation $\CC$. The mapping is
%%%%%%%%%%%%%%%%%%%%%%%%
\begin{subequations}\label{M2}
\begin{align}
& |\,\rho\rangle_{\B} \to |\,\rho\rangle_{\CC}=\sum_{\I \J} \rho_{\I \J} (a_1^\dag)^{I_1} \cdots (a_L^\dag)^{I_L} (c_1^\dag \hat{P})^{J_1} \cdots (c_L^\dag \hat{P})^{J_L} \,|0\rangle, \\
& \hat{L}_{\B} \to \hat{L}_{\CC}=\sum_{ij}\F_i(\a,\a^\dag) \, \F_j^\T (\hat{P}\c,\c^\dag \hat{P}).
\end{align}
\end{subequations}
%%%%%%%%%%%%%%%%%%%%%%%%
The LME and the expectation value of observable in representation $\CC$ are
%%%%%%%%%%%%%%%%%%%%%%%%
\begin{subequations}
\begin{align}
& \frac{d}{dt} |\,\rho\rangle_{\CC} = \hat{L}_{\CC} |\,\rho\rangle_{\CC}, \\
& \langle \O_{\a} \rangle = _{\CC} \! \langle \SS | \O_{\a} |\,\rho\rangle_{\CC},
\end{align}
\end{subequations}
%%%%%%%%%%%%%%%%%%%%%%%
where $_{\CC} \langle \SS |$ is defined as:
%%%%%%%%%%%%%%%%%%%%%%%%
\begin{equation}
_{\CC}\langle \SS |=\sum_{\S} \langle 0 |\, (\hat{P} c_L)^{S_L} \cdots (\hat{P} c_1)^{S_1} a_L^{S_L} \cdots a_1^{S_1}.
\end{equation}
%%%%%%%%%%%%%%%%%%%%%%%%
Combining the mappings in Eq.~(\ref{M1}) and Eq.~(\ref{M2}), we get the final mapping, i.e., Eq.~(\ref{Mapping}). The mapping process is schematically shown in Fig.~\ref{fig::8}.

%%%%%%%%%%%%%%%%%%%%%%%%%%%%%%%%%%%%%%%%%%%%%%%%%%%%%%%%%%%%%%%%%%%%%%%%%%%%%%%%%%%%%%%%%%%%%%%%%%%%%%%%%%%%%%%%%%%%%%%%
%%%%%%%%%%%%%%%%%%%%%%%%%%%%%%%%%%%%%%%%%%%%%%%%%%%%%%%%%%%%%%%%%%%%%%%%%%%%%%%%%%%%%%%%%%%%%%%%%%%%%%%%%%%%%%%%%%%%%%%%
\section{Diagonalize the model in representation $\CC$} \label{AP:2}

In this section we map our Liouvillian in Eq.~(\ref{LZ}) into representation $\CC$, get its BdG form in momentum space and diagonalize the Liouvillian.

%%%%%%%%%%%%%%%%%%%%%%%%%%%%%%%%%%%%%%%%%%%%%%%%%%%%%%%%%%%%%%%%%%%%%%%%%%%%%%%%%%%%%%%%%%%%%%%%%%%%%%%%%%%%%%%%%%%%%%%%
\subsection{ The derivation of $\hat{L}$} \label{AP:2:1}

Our Liouvillian $\L$ in Eq.~(\ref{LZ}) is mapped into $\hat{L}$ by the mapping~(\ref{MLZ}):
%%%%%%%%%%%%%%%%%%%%%%%%
\begin{equation}
\begin{split}
& \L(\cdot)=-i[H,\cdot]+(1-w)D^L(\cdot)+(1+w)D^R(\cdot) \\
& \to \ \hat{L}=\hat{H}+(1-w)\hat{D}^L+(1+w)\hat{D}^R.
\end{split}
\end{equation}
%%%%%%%%%%%%%%%%%%%%%%%%
Note that our matrix representation of creation and annihilation operator is real, thus we have $a^{\T}=a^\dag$, $c^{\T}= c^\dag$, $\hat{P}^{\T}=\hat{P}$. Then we get
%%%%%%%%%%%%%%%%%%%%%%%%
\begin{widetext}
\begin{equation}
-i[H,\cdot] \to \hat{H}  =  -iH(\a,\a^\dag) + i H^{\T} (\HP \c,\c^\dag \HP) = -iJ\sum_l (a^\dag_{l+1} a_l + a^\dag_l a_{l+1}) +iJ\sum_l (c^\dag_l c_{l+1} + c^\dag_{l+1} c_l ),
\end{equation}
\end{widetext}
%%%%%%%%%%%%%%%%%%%%%%%%
%%%%%%%%%%%%%%%%%%%%%%%%
\begin{widetext}
\begin{equation}
\begin{split}
&D^L(\cdot) \to \hat{D}^L  = \sum_l \Big( 2 A_l(\a,\a^\dag) A_l (\HP \c,\c^\dag \HP) -A^\dag_l(\a,\a^\dag) A_l(\a,\a^\dag) - A^\dag_l(\HP \c,\c^\dag \HP) A_l(\HP \c,\c^\dag \HP) \Big) \\
                         & = \sum_l  \Big( 2(\sqrt{\gamma_1} a_l^\dag + \sqrt{\gamma_2} a_{l+1} ) (\sqrt{\gamma_1} c^\dag_l \HP + \sqrt{\gamma_2} \HP c_{l+1}) - (\sqrt{\gamma_1}a_l +\sqrt{\gamma_2} a^\dag_{l+1}) (\sqrt{\gamma_1}a_l^\dag +\sqrt{\gamma_2} a_{l+1} ) \Big. \\
                         & \ \ \ \ \ \ \ \ \ \ \ \Big. -(\sqrt{\gamma_1} \HP c_l +\sqrt{\gamma_2} c^\dag_{l+1} \HP) (\sqrt{\gamma_1}c_l^\dag \HP +\sqrt{\gamma_2} \HP c_{l+1} ) \Big) \\
                         & = \sum_l \Big(-2\sqrt{\gamma_1 \gamma_2}\HP (a^\dag_l c_{l+1} + c^\dag_l a_{l+1}) + 2\gamma_1 \HP a^\dag_l c^\dag_l + 2\gamma_2 \HP c_{l+1} a_{l+1}-\sqrt{\gamma_1 \gamma_2}(a_l a_{l+1} + a^\dag_{l+1} a^\dag_l) \Big.\\
                         & \ \ \ \ \ \ \ \ \ \ \ \Big. + \sqrt{\gamma_1 \gamma_2}(c_l c_{l+1} + c^\dag_{l+1} c^\dag_l) -\gamma_2(a^\dag_{l+1} a_{l+1} + c^\dag_{l+1} c_{l+1}) -\gamma_1 (a_l a^\dag_l + c_l c^\dag_l)  \Big),
\end{split}
\end{equation}
\end{widetext}
%%%%%%%%%%%%%%%%%%%%%%%%
%%%%%%%%%%%%%%%%%%%%%%%%
\begin{widetext}
\begin{equation}
\begin{split}
&D^R(\cdot) \to \hat{D}^R  = \sum_l \Big( 2 A_l^\dag (\a,\a^\dag) A_l^\dag (\HP \c,\c^\dag \HP) -A_l(\a,\a^\dag) A_l^\dag(\a,\a^\dag) - A_l(\HP \c,\c^\dag \HP) A^\dag_l(\HP \c,\c^\dag \HP) \Big) \\
                         & = \sum_l  \Big( 2(\sqrt{\gamma_1} a_l + \sqrt{\gamma_2} a^\dag_{l+1} ) (\sqrt{\gamma_1} \HP c_l + \sqrt{\gamma_2} c^\dag_{l+1} \HP) - (\sqrt{\gamma_1}a^\dag_l +\sqrt{\gamma_2} a_{l+1}) (\sqrt{\gamma_1}a_l +\sqrt{\gamma_2} a^\dag_{l+1} ) \Big. \\
                         & \ \ \ \ \ \ \ \ \ \ \ \Big. -(\sqrt{\gamma_1} c^\dag_l \HP +\sqrt{\gamma_2} \HP c_{l+1}) (\sqrt{\gamma_1} \HP c_l +\sqrt{\gamma_2} c^\dag_{l+1} \HP ) \Big) \\
                         & = \sum_l \Big(-2\sqrt{\gamma_1 \gamma_2}\HP (a^\dag_{l+1} c_l + c^\dag_{l+1} a_l) + 2\gamma_1 \HP c_l a_l + 2\gamma_2 \HP a^\dag_{l+1} c^\dag_{l+1}+\sqrt{\gamma_1 \gamma_2}(a_l a_{l+1} + a^\dag_{l+1} a^\dag_l) \Big.\\
                         & \ \ \ \ \ \ \ \ \ \ \ \Big. - \sqrt{\gamma_1 \gamma_2}(c_l c_{l+1} + c^\dag_{l+1} c^\dag_l) -\gamma_2(a_{l+1} a^\dag_{l+1} + c_{l+1} c^\dag_{l+1}) -\gamma_1 (a^\dag_l a_l + c^\dag_l c_l)  \Big).
\end{split}
\end{equation}
\end{widetext}
%%%%%%%%%%%%%%%%%%%%%%%%
Due to $[\HP, \hat{L}]=0$, the state will keep its parity in the evolution governed by the Lindblad master equation. Therefore, $\HP$ can reduce to a constant $P$, which equals $1$ in even parity channel and $-1$ in odd parity channel. By Fourier transformation
%%%%%%%%%%%%%%%%%%%%%%%%
\begin{equation}
\begin{split}
& a^\dag_l =\sum_{k=-\pi}^{\pi} e^{-ikl} a^\dag_k,\ \ \ a_l=\sum_{k=-\pi}^{\pi} e^{ikl} a_k,\\
& c^\dag_l =\sum_{k=-\pi}^{\pi} e^{-ikl} c^\dag_k,\ \ \ c_l=\sum_{k=-\pi}^{\pi} e^{ikl} c_k,
\end{split}
\end{equation}
%%%%%%%%%%%%%%%%%%%%%%%%
we get $\hat{L}$ in BdG form as Eq.~(\ref{Lk0})$\sim$Eq.~(\ref{Lk1}).

%%%%%%%%%%%%%%%%%%%%%%%%%%%%%%%%%%%%%%%%%%%%%%%%%%%%%%%%%%%%%%%%%%%%%%%%%%%%%%%%%%%%%%%%%%%%%%%%%%%%%%%%%%%%%%%%%%%%%%%%
\subsection{ Diagonalization of $\hat{L}_k$} \label{AP:2:2}
We make a similarity transformation for $\hat{L}_k$ in Eq.~(\ref{Lk05}) by matrix $W$:
%%%%%%%%%%%%%%%%%%%%%%%%
\begin{equation}
\begin{split}
& \hat{L}_k  =(a^\dag_k \ c^\dag_k \ a_{-k} \ c_{-k}) \, W \, W^{-1} \, \L_k \, W \, W^{-1} \, (a_k \ c_k \ a^\dag_{-k} \ c^\dag_{-k} )^{\T} -4\gamma \\
& =(\zeta^{'}_1 (k) \ \zeta^{'}_2 (k) \ \zeta_3 (k) \ \zeta_4 (k))\  \Lambda \  (\zeta_1 (k) \ \zeta_2 (k) \ \zeta^{'}_3 (k) \ \zeta^{'}_4 (k))^{\T} -4\gamma \\
& =\lambda_1 (k) \zeta^{'}_1 (k) \zeta_1 (k) + \lambda_2 (k) \zeta^{'}_2 (k) \zeta_2 (k) \\
& + \lambda_3 (k) \zeta_3 (k) \zeta^{'}_3 (k)  +  \lambda_4 (k) \zeta_4 (k) \zeta^{'}_4 (k) -4\gamma,
\end{split}
\end{equation}
%%%%%%%%%%%%%%%%%%%%%%%
where
%%%%%%%%%%%%%%%%%%%%%%%%
\begin{equation}
\begin{split}
& (a^\dag_k \ c^\dag_k \ a_{-k} \ c_{-k}) \, W = (\zeta^{'}_1 (k) \ \zeta^{'}_2 (k) \ \zeta_3 (k) \ \zeta_4 (k)),\\
& W^{-1} \, (a_k \ c_k \ a^\dag_{-k} \ c^\dag_{-k} )^{\T}=(\zeta_1 (k) \ \zeta_2 (k) \ \zeta^{'}_3 (k) \ \zeta^{'}_4 (k))^{\T},
\end{split}
\end{equation}
%%%%%%%%%%%%%%%%%%%%%%%
and $\Lambda$ is a diagonal matrix given by
%%%%%%%%%%%%%%%%%%%%%%%%
\begin{equation}
\Lambda=W^{-1}\, \L_k \, W = \mathrm{diag}(\lambda_1 (k), \lambda_2 (k), \lambda_3 (k), \lambda_4 (k)).
\end{equation}
%%%%%%%%%%%%%%%%%%%%%%%
We write $W$ and $W^{-1}$ as
%%%%%%%%%%%%%%%%%%%%%%%%
\begin{equation}
W=(\vv_1 \ \vv_2 \ \vv_3 \ \vv_4),\ \ \ \ W^{-1}=\left( \begin{matrix} \uu_1^{\,t} \\  \uu_2^{\,t} \\  \uu_3^{\,t} \\  \uu_4^{\,t}  \end{matrix} \right),
\end{equation}
%%%%%%%%%%%%%%%%%%%%%%%
where the column vector $\vv_i$ and row vector $\uu_j^{\,t}$ satisfy $\uu_j^{\,t} \cdot \vv_i = \delta_{ij}$.
Then we have
%%%%%%%%%%%%%%%%%%%%%%%%
\begin{equation}~\label{NMM}
\begin{split}
&\zeta^{'}_1 (k)=(a^\dag_k \ c^\dag_k \ a_{-k} \ c_{-k}) \cdot \vv_1,\ \ \ \zeta^{'}_2 (k)=(a^\dag_k \ c^\dag_k \ a_{-k} \ c_{-k}) \cdot \vv_2,\\
&\zeta^{'}_3 (k)=(a_k \ c_k \ a^\dag_{-k} \ c^\dag_{-k}) \cdot \uu_3,\ \ \ \zeta^{'}_4 (k)=(a_k \ c_k \ a^\dag_{-k} \ c^\dag_{-k}) \cdot \uu_4,\\
&\zeta_1 (k)=(a_k \ c_k \ a^\dag_{-k} \ c^\dag_{-k}) \cdot \uu_1,\ \ \ \zeta_2 (k)=(a_k \ c_k \ a^\dag_{-k} \ c^\dag_{-k}) \cdot \uu_2,\\
&\zeta_3 (k)=(a^\dag_k \ c^\dag_k \ a_{-k} \ c_{-k}) \cdot \vv_3, \ \ \  \zeta_4 (k)=(a^\dag_k \ c^\dag_k \ a_{-k} \ c_{-k}) \cdot \vv_4.
\end{split}
\end{equation}
%%%%%%%%%%%%%%%%%%%%%%%
$\zeta^{'}_i(k)$ and $\zeta_j(k)$ hold anticommutation relations:
%%%%%%%%%%%%%%%%%%%%%%%%
\begin{equation}
\{\zeta^{'}_i (k),\zeta_j (k)\}=\delta_{ij},\ \ \ \{\zeta^{'}_i(k),\zeta^{'}_j(k)\}=\{\zeta_i(k),\zeta_j(k)\}=0
\end{equation}
%%%%%%%%%%%%%%%%%%%%%%%
Calculating the eigenvalues of Eq.~(\ref{Lk1}), we get the same values for both even and odd parity:
%%%%%%%%%%%%%%%%%%%%%%%%
\begin{equation}
\begin{split}
&\lambda_1 (k)=-2\gamma-2m_k,\ \ \ \lambda_2 (k)=-2\gamma+2m_k,\\
&\lambda_3 (k)=2\gamma-2m_k,\ \ \ \ \ \lambda_4 (k)=2\gamma+2m_k,
\end{split}
\end{equation}
%%%%%%%%%%%%%%%%%%%%%%%%
where $m_k$ is in Eq.~(\ref{mk}). Then $\L_k$ can be diagonalized as Eq.~(\ref{Lk+-}).

%%%%%%%%%%%%%%%%%%%%%%%%%%%%%%%%%%%%%%%%%%%%%%%%%%%%%%%%%%%%%%%%%%%%%%%%%%%%%%%%%%%%%%%%%%%%%%%%%%%%%%%%%%%%%%%%%%%%%%%%
%%%%%%%%%%%%%%%%%%%%%%%%%%%%%%%%%%%%%%%%%%%%%%%%%%%%%%%%%%%%%%%%%%%%%%%%%%%%%%%%%%%%%%%%%%%%%%%%%%%%%%%%%%%%%%%%%%%%%%%%
\section{Exactly solution of the model when $w=0$} \label{AP:3}
In this section, we exactly solve our model both in even and odd channels. We show steady state and all the excited states of the open system. In addition, we prove that the odd parity states have no contribution on observations with even fermionic operators. Last, we calculate the correlation functions of steady state and local-quantum-jump states beyond the steady state.

%%%%%%%%%%%%%%%%%%%%%%%%%%%%%%%%%%%%%%%%%%%%%%%%%%%%%%%%%%%%%%%%%%%%%%%%%%%%%%%%%%%%%%%%%%%%%%%%%%%%%%%%%%%%%%%%%%%%%%%%
\subsection{ All the eigenstates of $\hat{L}$} \label{AP:3:1}
First, we diagonalize $\hat{L}_k$ in even channel ($P=1$). Then normal master modes are show in Eq.~(\ref{NMM}). The vectors $\vv$ and $\uu$ can be solved as
%%%%%%%%%%%%%%%%%%%%%%%%
\begin{equation}
\begin{split}
& \vv_1=\frac{1}{2} \Big( -1-iJ\cos k/m_k,\, -2\sqrt{\gamma_1\gamma_2} \cos k/m_k,\\
&\ \ \ \ \ \ \ \ \ \ \ -2\sqrt{\gamma_1\gamma_2} \cos k/m_k,\,  1+iJ\cos k/m_k   \Big)^{\T} \\
& \vv_2=\frac{1}{2} \Big( -1+iJ\cos k/m_k,\, 2\sqrt{\gamma_1\gamma_2} \cos k/m_k,\\
&\ \ \ \ \ \ \ \ \  \ \ 2\sqrt{\gamma_1\gamma_2} \cos k/m_k,\,  1-iJ\cos k/m_k   \Big)^{\T} \\
& \vv_3=\frac{1}{2} \Big(1,\, \frac{-iJ+m_k/\cos k}{2\sqrt{\gamma_1\gamma_2}},\, \frac{iJ-m_k/\cos k}{2\sqrt{\gamma_1\gamma_2}},\, 1  \Big)^{\T} \\
& \vv_4=\frac{1}{2} \Big(1,\, \frac{-iJ-m_k/\cos k}{2\sqrt{\gamma_1\gamma_2}},\, \frac{iJ+m_k/\cos k}{2\sqrt{\gamma_1\gamma_2}},\, 1  \Big)^{\T} \\
& \uu_1=\frac{1}{2} \Big(-1,\, \frac{iJ-m_k/\cos k}{2\sqrt{\gamma_1\gamma_2}},\, \frac{iJ-m_k/\cos k}{2\sqrt{\gamma_1\gamma_2}},\, 1  \Big)^{\T} \\
& \uu_2=\frac{1}{2} \Big(-1,\, \frac{iJ+m_k/\cos k}{2\sqrt{\gamma_1\gamma_2}},\, \frac{iJ+m_k/\cos k}{2\sqrt{\gamma_1\gamma_2}},\, 1  \Big)^{\T} \\
& \uu_3=\frac{1}{2} \Big( 1+iJ\cos k/m_k,\, 2\sqrt{\gamma_1\gamma_2} \cos k/m_k,\\
&\ \ \ \ \ \ \ \ \ \ \  -2\sqrt{\gamma_1\gamma_2} \cos k/m_k,\,  1+iJ\cos k/m_k   \Big)^{\T} \\
& \uu_4=\frac{1}{2} \Big( 1-iJ\cos k/m_k,\, -2\sqrt{\gamma_1\gamma_2} \cos k/m_k,\\
&\ \ \ \ \ \ \ \ \ \ \  2\sqrt{\gamma_1\gamma_2} \cos k/m_k,\,  1-iJ\cos k/m_k   \Big)^{\T}.
\end{split}
\end{equation}
%%%%%%%%%%%%%%%%%%%%%%%
We make an ansatz for steady state $|\Omega\rangle$ as
%%%%%%%%%%%%%%%%%%%%%%%%
\begin{equation}
|\Omega\rangle= \prod_{k=0}^{\pi} (z_1 + z_2 a^\dag_k c^\dag_{-k})(z_3 + z_4 a^\dag_{-k} c^\dag_k) |0\rangle.
\end{equation}
%%%%%%%%%%%%%%%%%%%%%%%%
Solving the steady state equations: $\zeta_i |\Omega\rangle =0 $ for $i=1 \sim 4$, we get $z_1=z_2$ and $z_3=z_4$. Therefore, the solution of steady state (the Eq.~(16) in the main text) is given by
%%%%%%%%%%%%%%%%%%%%%%%%%
\begin{equation}
|\Omega\rangle=\frac{1}{\N}\prod_{k=-\pi}^{\pi} (1+a^\dag_k c^\dag_{-k}) |0\rangle.
\end{equation}
%%%%%%%%%%%%%%%%%%%%%%%%%
By using $\Tr(\rho_{s})=1$, we get the normalization factor $\N$ as
%%%%%%%%%%%%%%%%%%%%%%%%%
\begin{equation}
\N=_{\CC}\! \langle \S_0 | \prod_{k=-\pi}^{\pi} (1+a^\dag_k c^\dag_{-k}) |0\rangle = 2^L,
\end{equation}
%%%%%%%%%%%%%%%%%%%%%%%%%
where $L$ is the length of the chain. The details of $\N=2^L$ is given in subsection 4. In addition, we get steady state in real space given by
%%%%%%%%%%%%%%%%%%%%%%%%%
\begin{equation}
\begin{split}
 &|\Omega\rangle=\frac{1}{\N} \exp \Big(\sum_{k=-\pi}^{\pi} a^\dag_k c^\dag_{-k} \Big) |0\rangle = \frac{1}{\N} \exp \Big(\sum_{l=1}^{L} a^\dag_l c^\dag_l \Big)|0\rangle \\
 & = \frac{1}{\N}\prod_{l=1}^{L} (1+a^\dag_l c^\dag_l) |0\rangle.
 \end{split}
\end{equation}
%%%%%%%%%%%%%%%%%%%%%%%%%
Under the parity constraint, valid eigenstates in even parity channel are $|\Omega\rangle$,  $\zeta^{'}_{\alpha_1} (k_i) \zeta^{'}_{\alpha_2} (k_j) |\Omega\rangle$,  $\zeta^{'}_{\alpha_1} (k_i) \zeta^{'}_{\alpha_2} (k_j) \zeta^{'}_{\alpha_3} (k_m) \zeta^{'}_{\alpha_4} (k_n) |\Omega\rangle $,  $\cdots$

Secondly, we diagonalize $\hat{L}_k$ in the odd channel ($P=-1$). The process of diagonalization is the same as it in the even channel, however, the eigenvectors $\vv$ and $\uu$ of odd channel are different from them in even channel. We mark the eigenvectors and normal master modes of the odd channel with '$*$':
%%%%%%%%%%%%%%%%%%%%%%%%
\begin{equation}
\begin{split}
&\zeta^{'}_{1*} (k)=(a^\dag_k \ c^\dag_k \ a_{-k} \ c_{-k}) \cdot \vv_{1*},\ \ \ \zeta^{'}_{2*} (k)=(a^\dag_k \ c^\dag_k \ a_{-k} \ c_{-k}) \cdot \vv_{2*},\\
& \zeta^{'}_{3*} (k)=(a_k \ c_k \ a^\dag_{-k} \ c^\dag_{-k}) \cdot \uu_{3*},\ \ \ \zeta^{'}_{4*} (k)=(a_k \ c_k \ a^\dag_{-k} \ c^\dag_{-k}) \cdot \uu_{4*},\\
&\zeta_{1*} (k)=(a_k \ c_k \ a^\dag_{-k} \ c^\dag_{-k}) \cdot \uu_{1*},\ \ \ \zeta_{2*} (k)=(a_k \ c_k \ a^\dag_{-k} \ c^\dag_{-k}) \cdot \uu_{2*},\\
& \zeta_{3*} (k)=(a^\dag_k \ c^\dag_k \ a_{-k} \ c_{-k}) \cdot \vv_{3*}, \ \ \  \zeta_{4*} (k)=(a^\dag_k \ c^\dag_k \ a_{-k} \ c_{-k}) \cdot \vv_{4*},
\end{split}
\end{equation}
%%%%%%%%%%%%%%%%%%%%%%
where
%%%%%%%%%%%%%%%%%%%%%%%%
\begin{equation}
\begin{split}
& \vv_{1*}=\frac{1}{2} \Big( 1+iJ\cos k/m_k,\, -2\sqrt{\gamma_1\gamma_2} \cos k/m_k,\\
&\ \ \ \ \ \ \ \ \ \ \  2\sqrt{\gamma_1\gamma_2} \cos k/m_k,\,  1+iJ\cos k/m_k   \Big)^{\T} \\
& \vv_{2*}=\frac{1}{2} \Big( 1-iJ\cos k/m_k,\, 2\sqrt{\gamma_1\gamma_2} \cos k/m_k,\\
&\ \ \ \ \ \ \ \ \ \ \  -2\sqrt{\gamma_1\gamma_2} \cos k/m_k,\,  1-iJ\cos k/m_k   \Big)^{\T} \\
& \vv_{3*}=\frac{1}{2} \Big(-1,\, \frac{-iJ+m_k/\cos k}{2\sqrt{\gamma_1\gamma_2}},\, \frac{-iJ+m_k/\cos k}{2\sqrt{\gamma_1\gamma_2}},\, 1  \Big)^{\T} \\
& \vv_{4*}=\frac{1}{2} \Big(-1,\, \frac{-iJ-m_k/\cos k}{2\sqrt{\gamma_1\gamma_2}},\, \frac{-iJ-m_k/\cos k}{2\sqrt{\gamma_1\gamma_2}},\, 1  \Big)^{\T} \\
& \uu_{1*}=\frac{1}{2} \Big(1,\, \frac{iJ-m_k/\cos k}{2\sqrt{\gamma_1\gamma_2}},\, \frac{-iJ+m_k/\cos k}{2\sqrt{\gamma_1\gamma_2}},\, 1  \Big)^{\T} \\
& \uu_{2*}=\frac{1}{2} \Big(1,\, \frac{iJ+m_k/\cos k}{2\sqrt{\gamma_1\gamma_2}},\, \frac{-iJ-m_k/\cos k}{2\sqrt{\gamma_1\gamma_2}},\, 1  \Big)^{\T} \\
& \uu_{3*}=\frac{1}{2} \Big( -1-iJ\cos k/m_k,\, 2\sqrt{\gamma_1\gamma_2} \cos k/m_k,\\
&\ \ \ \ \ \ \ \ \ \ \  2\sqrt{\gamma_1\gamma_2} \cos k/m_k,\,  1+iJ\cos k/m_k   \Big)^{\T} \\
& \uu_{4*}=\frac{1}{2} \Big( -1+iJ\cos k/m_k,\, -2\sqrt{\gamma_1\gamma_2} \cos k/m_k,\\
&\ \ \ \ \ \ \ \ \ \ \  -2\sqrt{\gamma_1\gamma_2} \cos k/m_k,\,  1-iJ\cos k/m_k   \Big)^{\T}.
\end{split}
\end{equation}
%%%%%%%%%%%%%%%%%%%%%%%
Solving the equation, $\zeta_{i*} |\Omega * \rangle =0 $ for $i=1 \sim 4$, we get
%%%%%%%%%%%%%%%%%%%%%%%%%
\begin{equation}
|\Omega * \rangle = \frac{1}{\N}\prod_{k=-\pi}^{\pi} (1-a^\dag_k c^\dag_{-k}) |0\rangle = \frac{1}{\N}\prod_{l=1}^{L} (1-a^\dag_l c^\dag_l) |0\rangle.
\end{equation}
%%%%%%%%%%%%%%%%%%%%%%%%%
Note that $|\Omega * \rangle$ is even parity ($\hat{P}|\Omega * \rangle=+1|\Omega * \rangle$). Therefore, the valid eigenstates in odd parity channel are the states with odd numbers of excitations on the $|\Omega * \rangle$, i.e. $\zeta^{'}_{\alpha_1 *} (k_i) |\Omega * \rangle$,  $\zeta^{'}_{\alpha_1 *} (k_i) \zeta^{'}_{\alpha_2 *} (k_j) \zeta^{'}_{\alpha_3 *} (k_m) |\Omega *\rangle $, $\cdots$

In summary, the full eigenstates of $\hat{L}$ are
%%%%%%%%%%%%%%%%%%%%%%%%
\begin{equation} \label{FullModes}
\begin{split}
&\text{Steady state:} \ \ \ \ \ \ \ \ \ \ \ \ \ \ \ \ \ \ \ \ \ |\Omega\rangle \\
&\text{Single excitation:} \ \ \ \ \ \ \ \ \ \ \ \ \ \zeta^{'}_{\alpha_1 *} (k_i) \ |\Omega * \rangle \\
&\text{Double excitation:} \ \ \ \ \ \ \ \ \ \ \ \zeta^{'}_{\alpha_1} (k_i) \zeta^{'}_{\alpha_2} (k_j) \ |\Omega\rangle \\
&\text{Triple excitation:} \ \ \ \ \ \ \ \ \ \ \ \ \ \zeta^{'}_{\alpha_1 *} (k_i) \zeta^{'}_{\alpha_2 *} (k_j) \zeta^{'}_{\alpha_3 *} (k_m) \ |\Omega *\rangle \\
&\text{Quadruple excitation:} \ \ \ \ \ \zeta^{'}_{\alpha_1} (k_i) \zeta^{'}_{\alpha_2} (k_j) \zeta^{'}_{\alpha_3} (k_m) \zeta^{'}_{\alpha_4} (k_n) \ |\Omega\rangle \\
& \ \ \ \ \ \ \ \ \cdots
\end{split}
\end{equation}
%%%%%%%%%%%%%%%%%%%%%%%

%%%%%%%%%%%%%%%%%%%%%%%%%%%%%%%%%%%%%%%%%%%%%%%%%%%%%%%%%%%%%%%%%%%%%%%%%%%%%%%%%%%%%%%%%%%%%%%%%%%%%%%%%%%%%%%%%%%%%%%%
\subsection{  Flat band condition} \label{AP:3:2}
When the condition $J^2 = 4\gamma_1 \gamma_2$ is satisfied, Liouvillian flat band occurs. We have $\lambda_1 = \lambda_2 = -2\gamma$, $\lambda_3 = \lambda_4 = 2\gamma$ and $m_k=0$, which leads to divergence of eigenvectors $\vv_{1}$, $\vv_{2}$, $\uu_{3}$, $\uu_{4}$, $\vv_{1*}$, $\vv_{2*}$, $\uu_{3*}$ and $\uu_{4*}$. This indicates the exceptional point of $\hat{L}$. However, we can eliminate divergence by summing of these eigenvectors. Setting $J=2\sqrt{\gamma_1 \gamma_2}$, we can get the normal master modes in even parity
%%%%%%%%%%%%%%%%%%%%%%%%
\begin{equation}
\begin{split}
& \zeta^{'}_A (k) = (a^\dag_k \ c^\dag_k \ a_{-k} \ c_{-k} ) \cdot (\vv_1 + \vv_2) = -a^\dag_k + c_{-k}   \\
& \zeta_A (k) = (a_k\ c_k\ a^\dag_{-k}\ c^\dag_{-k}) \cdot (\uu_1 + \uu_2) /2 \\
& = \frac{1}{2}(-a_k+ic_k+ia^\dag_{-k}+c^\dag_{-k}) \\
&\zeta^{'}_B (k) = a_k\ c_k\ a^\dag_{-k}\ c^\dag_{-k}) \cdot (\uu_3 + \uu_4) = a_k + c^\dag_{-k}   \\
& \zeta_B (k) =(a^\dag_k \ c^\dag_k \ a_{-k} \ c_{-k} ) \cdot (\vv_3 + \vv_4) /2 \\
& = \frac{1}{2}(a^\dag_k-ic^\dag_k+ia_{-k}+c_{-k}),
\end{split}
\end{equation}
%%%%%%%%%%%%%%%%%%%%%%%
and in odd parity
%%%%%%%%%%%%%%%%%%%%%%%%
\begin{equation}
\begin{split}
& \zeta^{'}_{A*} (k) = (a^\dag_k \ c^\dag_k \ a_{-k} \ c_{-k} ) \cdot (\vv_{1*} + \vv_{2*}) = a^\dag_k + c_{-k}   \\
& \zeta_{A*} (k) = (a_k\ c_k\ a^\dag_{-k}\ c^\dag_{-k}) \cdot (\uu_{1*} + \uu_{2*}) /2 \\
& = \frac{1}{2}(a_k+ic_k-ia^\dag_{-k}+c^\dag_{-k}) \\
&\zeta^{'}_{B*} (k) = a_k\ c_k\ a^\dag_{-k}\ c^\dag_{-k}) \cdot (\uu_{3*} + \uu_{4*}) = -a_k + c^\dag_{-k}   \\
& \zeta_{B*} (k) =(a^\dag_k \ c^\dag_k \ a_{-k} \ c_{-k} ) \cdot (\vv_{3*} + \vv_{4*}) /2 \\
& = \frac{1}{2}(-a^\dag_k-ic^\dag_k-ia_{-k}+c_{-k}).
\end{split}
\end{equation}
%%%%%%%%%%%%%%%%%%%%%%%

%%%%%%%%%%%%%%%%%%%%%%%%%%%%%%%%%%%%%%%%%%%%%%%%%%%%%%%%%%%%%%%%%%%%%%%%%%%%%%%%%%%%%%%%%%%%%%%%%%%%%%%%%%%%%%%%%%%%%%%%
\subsection{ Ineffectiveness of odd parity } \label{AP:3:3}

Given an arbitrary state $|\,\rho\rangle$, it can be decomposed into even and odd eigenstate of $\hat{L}$:
%%%%%%%%%%%%%%%%%%%%%%%%%
\begin{equation}
|\,\rho\rangle = \Big( \sum_i C^{e}_i \,| i\rangle_e  \Big) + \Big( \sum_j C^{o}_j \,| j\rangle_o  \Big),
\end{equation}
%%%%%%%%%%%%%%%%%%%%%%%%%
where $| i\rangle_e$  and $| j\rangle_o$ represents even and odd parity state in Eq.~(\ref{FullModes}). The expectation value of observation $\hat{O}$ is
%%%%%%%%%%%%%%%%%%%%%%%%%
\begin{equation}
{_\CC} \langle \S_0| \hat{O} |\,\rho\rangle = \Big( \sum_i C^{e}_i\, {_\CC} \langle \S_0| \hat{O} | i\rangle_e  \Big) + \Big( \sum_j C^{o}_j \, {_\CC} \langle \S_0| \hat{O} | j\rangle_o  \Big).
\end{equation}
%%%%%%%%%%%%%%%%%%%%%%%%%
When $\hat{O}$ has even fermionic operators, we have ${_\CC}\langle \S_0| \hat{O} | j\rangle_o = 0$. When $\hat{O}$ has odd fermionic operators, we have ${_\CC}\langle \S_0| \hat{O} | i\rangle_e = 0$. Usually, in pure fermionic system, fermionic operators appear in pairs, so the odd parity part of $\hat{L}$ does not influence the expectation value of observation.

%%%%%%%%%%%%%%%%%%%%%%%%%%%%%%%%%%%%%%%%%%%%%%%%%%%%%%%%%%%%%%%%%%%%%%%%%%%%%%%%%%%%%%%%%%%%%%%%%%%%%%%%%%%%%%%%%%%%%%%%
\subsection{ Correlation functions of steady state and quantum jump states} \label{AP:3:4}

Firstly, we show the details for the calculation of normalization factor $\N$:
%%%%%%%%%%%%%%%%%%%%%%%%
\begin{equation}
\begin{split}
&\N = _{\CC}\! \langle \S_0 | \prod_{l=1}^{L} (1+a^\dag_l c^\dag_l) |0\rangle \\
   &=  \sum_{\S} \langle 0|   (\hat{P} c_L)^{S_L} \cdots (\hat{P} c_1)^{S_1} a_L^{S_L} \cdots a_1^{S_1}  \\
   &\ \ \ \ \ \ \ \ \ \ \  \ (1+a^\dag_1 c^\dag_1)\cdots (1+a^\dag_L c^\dag_L) |0\rangle \\
   &= \sum_{\S} \langle 0| (\hat{P} c_L a_L)^{S_L} \cdots (\hat{P} c_1 a_1)^{S_1}\  (1+a^\dag_1 c^\dag_1)\cdots (1+a^\dag_L c^\dag_L) |0\rangle \\
   &= \langle 0| (1+\hat{P} c_L a_L) \cdots (1+\hat{P} c_1 a_1)\  (1+a^\dag_1 c^\dag_1)\cdots (1+a^\dag_L c^\dag_L) |0\rangle \\
   &=  \langle 0| \prod_{l=1}^{L} \Big( (1+\hat{P} c_l a_l) (1+a^\dag_l c^\dag_l) \Big) |0\rangle \\
   &= 2^L.
\end{split}
\end{equation}
%%%%%%%%%%%%%%%%%%%%%%%

Secondly, we show the particle number distribution of the steady state $n^s_j$
%%%%%%%%%%%%%%%%%%%%%%%%%
\begin{equation}
\begin{split}
&n^s_j = _{\CC}\! \langle \S_0|  a^\dag_j a_j |\Omega\rangle \\
    &=\frac{1}{\N} \sum_{\S} \langle 0| (\hat{P} c_L)^{S_L} \cdots (\hat{P} c_1)^{S_1}  a_L^{S_L} \cdots  a_1^{S_1} \,  a^\dag_j a_j \\
    &\ \ \ \ \ \ \ \ \ \ \ \ \ \  \ \ \ (1+a^\dag_1 c^\dag_1)\cdots (1+a^\dag_L c^\dag_L) |0\rangle  \\
    &= \frac{2^{L-1}}{\N} \langle 0| (1+\hat{P} c_j a_j) a^\dag_j a_j (1+a^\dag_j c^\dag_j) |0\rangle  \\
    &=\frac{1}{2}
\end{split}
\end{equation}
%%%%%%%%%%%%%%%%%%%%%%%%%
The other correlation functions of steady state can be calculated by the same method. The results are
%%%%%%%%%%%%%%%%%%%%%%%%%
\begin{equation}
G^s_{j_1, j_2}=0\, (j_1 \ne j_2), \ \ D^s_{j_1,j_2}=0,\ \ D^{s*}_{j_1,j_2}=0.
\end{equation}
%%%%%%%%%%%%%%%%%%%%%%%%%

Thirdly, we focus on a state from a quantum jump on the site $l$ of the steady state. We denote this state as $|\phi^l\rangle$:
%%%%%%%%%%%%%%%%%%%%%%%%%
\begin{equation}
|\phi^l\rangle:= \frac{a^\dag_l \rho_s a_l }{\Tr (a^\dag_l \rho_s a_l)} =  \frac{a^\dag_l c^\dag_l |\Omega\rangle}{ {_\CC}\! \langle \S_0| a^\dag_l c^\dag_l |\Omega\rangle}.
\end{equation}
%%%%%%%%%%%%%%%%%%%%%%%%%
The particle number on site $j$ of $|\phi^l\rangle$, denoted as $n^l_j$:
%%%%%%%%%%%%%%%%%%%%%%%%%
\begin{equation}
\begin{split}
& n^l_{j=l} =G^l_{l, l}= _{\CC}\! \langle \S_0|  a^\dag_l a_l |\phi^l\rangle \\
    &= \frac{\langle 0| (1+\hat{P} c_l a_l)\, a^\dag_l a_l\,  a^\dag_l c^\dag_l  (1+a^\dag_l c^\dag_l) |0\rangle }{\langle 0| (1+\hat{P} c_l a_l) a^\dag_l c^\dag_l  (1+a^\dag_l c^\dag_l) |0\rangle } \\
    &=1.
\end{split}
\end{equation}
%%%%%%%%%%%%%%%%%%%%%%%%
%%%%%%%%%%%%%%%%%%%%%%%%%
\begin{equation}
\begin{split}
& n^l_{j\ne l} =G^l_{j, j}= _{\CC}\! \langle \S_0|  a^\dag_j a_j |\phi^l\rangle \ (j\ne l)\\
    &= \frac{\langle 0| (1+\hat{P} c_l a_l) a^\dag_l c^\dag_l (1+a^\dag_l c^\dag_l)\, (1+\hat{P} c_j a_j) a^\dag_j c^\dag_j (1+a^\dag_j c^\dag_j) |0\rangle }{\langle 0| (1+\hat{P} c_l a_l) a^\dag_l c^\dag_l (1+a^\dag_l c^\dag_l)\, (1+\hat{P} c_j a_j) (1+a^\dag_j c^\dag_j) |0\rangle} \\
    &=\frac{1}{2}.
\end{split}
\end{equation}
%%%%%%%%%%%%%%%%%%%%%%%%
By the same way, we get the other correlation functions of $|\phi^l\rangle$. The results are
%%%%%%%%%%%%%%%%%%%%%%%%%
\begin{equation}
G^l_{j_1, j_2}=0\, (j_1 \ne j_2), \ \ D^l_{j_1,j_2}=0,\ \ D^{l*}_{j_1,j_2}=0.
\end{equation}
%%%%%%%%%%%%%%%%%%%%%%%%%

%%%%%%%%%%%%%%%%%%%%%%%%%%%%%%%%%%%%%%%%%%%%%%%%%%%%%%%%%%%%%%%%%%%%%%%%%%%%%%%%%%%%%%%%%%%%%%%%%%%%%%%%%%%%%%%%%%%%%%%%
%%%%%%%%%%%%%%%%%%%%%%%%%%%%%%%%%%%%%%%%%%%%%%%%%%%%%%%%%%%%%%%%%%%%%%%%%%%%%%%%%%%%%%%%%%%%%%%%%%%%%%%%%%%%%%%%%%%%%%%%
\section{Evolution equations of correlation functions in real space} \label{AP:4}

The evolution equation of the expectation value of operator $\O$ in the open system is
%%%%%%%%%%%%%%%%%%%%%%%%%
\begin{equation}~\label{RO0}
\frac{d}{dt} \Tr (\O \rho(t)) = \Tr (\O \frac{d}{dt} \rho) = \Tr (\O \L(\rho)).
\end{equation}
%%%%%%%%%%%%%%%%%%%%%%%%
By considering the Liouvillian $\L$ in Eq.~(\ref{LZ}), the equation becomes
%%%%%%%%%%%%%%%%%%%%%%%%%
\begin{equation}~\label{RO1}
\begin{split}
& \frac{d}{dt} \Tr (\O \rho(t)) = -i \Tr (\O [H,\rho]) \\
& +(1-w) \Tr (\O D^L (\rho)) + (1+w) \Tr (\O D^R (\rho)).
\end{split}
\end{equation}
%%%%%%%%%%%%%%%%%%%%%%%%
Using the relation $\Tr(ABC)=\Tr(CAB)$, we have
%%%%%%%%%%%%%%%%%%%%%%%%%
\begin{equation}~\label{RO2}
\Tr(\O [H,\rho])=\Tr([\O,H] \, \rho) = J \sum_{l} \Tr( [\O, a^\dag_{l+1} a_l + a^\dag_l a_{l+1}  ] \, \rho),
\end{equation}
%%%%%%%%%%%%%%%%%%%%%%%%
%%%%%%%%%%%%%%%%%%%%%%%%%
\begin{equation}~\label{RO3}
\begin{split}
&\Tr(\O D^L(\rho)) \\
&= \sum_l \Big( \Tr (\O 2 A_l \, \rho \, A^\dag_l) - \Tr (\O A^\dag_l A_l \, \rho  ) - \Tr (\O \, \rho \, A^\dag_l A_l)   \Big) \\
&= \sum_l \Big(  \Tr ([A^\dag_l , \O] A_l \, \rho ) +   \Tr (A^\dag_l [\O , A_l] \, \rho )  \Big),\\
&\ \\
&\Tr(\O D^R(\rho)) \\
&= \sum_l \Big( \Tr (\O 2 A^\dag_l \, \rho \, A_l) - \Tr (\O A_l A^\dag_l \, \rho  ) - \Tr (\O \, \rho \, A_l A^\dag_l)   \Big) \\
&= \sum_l \Big(  \Tr ([A_l , \O] A^\dag_l \, \rho ) +   \Tr (A_l [\O , A^\dag_l] \, \rho )  \Big).
\end{split}
\end{equation}
%%%%%%%%%%%%%%%%%%%%%%%%
Substituting $\O = a^\dag_{l_1} a_{l_2}$, $\O = a_{l_1} a_{l_2}$ and $\O = a^\dag_{l_2} a^\dag_{l_1}$ into Eq.(\ref{RO0}) $\sim$ Eq.(\ref{RO3}), we get the evolution equations of $G_{l_1 , l_2}$, $D_{l_1 , l_2}$ and $D^{*}_{l_1 , l_2}$, respectively. Namely, the evolution equations of correlation functions in real space are
%%%%%%%%%%%%%%%%%%%%%%%%%
\begin{widetext}
\begin{equation}~\label{dtG}
\begin{split}
&\frac{d}{dt} G_{l_1,l_2} = -4 \gamma G_{l_1,l_2} + iJ(G_{l_1-1,l_2} + G_{l_1+1,l_2} - G_{l_1,l_2-1} - G_{l_1,l_2+1})+ 2\big[\gamma + w (\gamma_2 -\gamma_1)\big]\, \delta_{l_1,l_2}\\
&+\sqrt{\gamma_1 \gamma_2}\, (-D_{l_1-1,l_2} - D_{l_1+1,l_2} + D_{l_2,l_1-1} + D_{l_2,l_1+1} ) + \sqrt{\gamma_1 \gamma_2}\, (D^{*}_{l_1,l_2-1} + D^{*}_{l_1,l_2+1} - D^{*}_{l_2-1,l_1} - D^{*}_{l_2+1,l_1} )\\
&\ \\
&\frac{d}{dt} D_{l_1,l_2} = +2\sqrt{\gamma_1 \gamma_2} (-G_{l_1-1,l_2} - G_{l_1+1,l_2} + G_{l_2-1,l_1} + G_{l_2+1,l_1} )  + 2 w \sqrt{\gamma_1 \gamma_2} \big( \delta_{l_1,l_2-1} -\delta_{l_2,l_1-1} \big)\\
&-4 \gamma D_{l_1,l_2} - iJ/2\,(D_{l_1-1,l_2} + D_{l_1+1,l_2} + D_{l_1,l_2-1} + D_{l_1,l_2+1})+ iJ/2\,(D_{l_2,l_1-1} + D_{l_2,l_1+1} + D_{l_2-1,l_1} + D_{l_2+1,l_1})\\
&\ \\
&\frac{d}{dt} D^{*}_{l_1,l_2} = +2\sqrt{\gamma_1 \gamma_2} (-G_{l_2,l_1-1} - G_{l_2,l_1+1} + G_{l_1,l_2-1} + G_{l_1,l_2+1} )  + 2 w \sqrt{\gamma_1 \gamma_2} \big( \delta_{l_1,l_2-1} -\delta_{l_2,l_1-1} \big)\\
&-4 \gamma D^{*}_{l_1,l_2} + iJ/2\,(D^{*}_{l_1-1,l_2} + D^{*}_{l_1+1,l_2} + D^{*}_{l_1,l_2-1} + D^{*}_{l_1,l_2+1})-iJ/2\,(D^{*}_{l_2,l_1-1} + D^{*}_{l_2,l_1+1} + D^{*}_{l_2-1,l_1} + D^{*}_{l_2+1,l_1}).
\end{split}
\end{equation}
\end{widetext}
%%%%%%%%%%%%%%%%%%%%%%%%%
%%%%%%%%%%%%%%%%%%%%%%%%%%%%%%%%%%%%%%%%%%%%%%%%%%%%%%%%%%%%%%%%%%%%%%%%%%%%%%%%%%%%%%%%%%%%%%%%%%%%%%%%%%%%%%%%%%%%%%%%

%%%%%%%%%%%%%%%%%%%%%%%%%%%%%%%%%%%%%%%%%%%%%%%%%%%%%%%%%%%%%%%%%%%%%%%%%%%%%%%%%%%%%%%%%%%%%%%%%%%%%%%%%%%%%%%%%%%%%%%%
%%%%%%%%%%%%%%%%%%%%%%%%%%%%%%%%%%%%%%%%%%%%%%%%%%%%%%%%%%%%%%%%%%%%%%%%%%%%%%%%%%%%%%%%%%%%%%%%%%%%%%%%%%%%%%%%%%%%%%%%
\section{The relationship between the damping-matrix spectra and the Liouvillian spectra } \label{AP:5}

In this section, we demonstrate that for a real physical process with closed evolution equations of correlation functions the damping-matrix spectra are the subset of the Liouvillian spectra.

The general form of closed evolution equations of correlation functions is
%%%%%%%%%%%%%%%%%%%%%%%%
\begin{equation}
\frac{d}{dt} \Psi = \X \, \Psi + V,
\end{equation}
%%%%%%%%%%%%%%%%%%%%%%%%
where $\X$ is the damping matrix, $\Psi$ is the vector of correlation functions, for example, $\Psi$ is taken as $(G_{k_1,k_2},G_{-k_2,-k_1},D_{k_2,-k_1},D^*_{k_1,-k_2})^\T$ in our model. The vector $V$ induces the correlation function vector of steady state $\Psi^S$ as $\Psi^S = - \X^{-1} V$. By deducting $\Psi^S$, we have
%%%%%%%%%%%%%%%%%%%%%%%%
\begin{equation}~\label{DE}
\frac{d}{dt} (\Psi(t) -\Psi^S ) = \X \, (\Psi(t) -\Psi^S ).
\end{equation}
%%%%%%%%%%%%%%%%%%%%%%%%
If the correlation function vector $\Psi^\Gamma$ is governed by the eigen equation of damping matrix, we have
%%%%%%%%%%%%%%%%%%%%%%%%
\begin{equation}
\X \, (\Psi^\Gamma (t) - \Psi^S) = \Gamma \,  (\Psi^\Gamma (t) - \Psi^S),
\end{equation}
%%%%%%%%%%%%%%%%%%%%%%%%
where $\Gamma$ is the eigenvalue of $\X$. The equation in the initial time is
%%%%%%%%%%%%%%%%%%%%%%%%
\begin{equation}
\X \, (\Psi^\Gamma (0) - \Psi^S) = \Gamma \,  (\Psi^\Gamma (0) - \Psi^S),
\end{equation}
%%%%%%%%%%%%%%%%%%%%%%%%
Then from Eq.~(\ref{DE}), we obtain
%%%%%%%%%%%%%%%%%%%%%%%%
\begin{equation}~\label{Gt}
\Psi^\Gamma (t) - \Psi^S = e^{\Gamma t} \, (\Psi^\Gamma (0) - \Psi^S).
\end{equation}
%%%%%%%%%%%%%%%%%%%%%%%%
If $\Psi^\Gamma$ is in a real physical process, we will have
%%%%%%%%%%%%%%%%%%%%%%%%
\begin{subequations}~\label{RE}
\begin{align}
& \Psi^\Gamma (t) =\, _{\CC} \langle \SS |\, \hat{\Psi}\,  e^{\hat{L}_{\CC}\, t}  |\,\rho (0) \rangle_{\CC},\\
& \Psi^S =\, _{\CC} \langle \SS |\, \hat{\Psi}\,  e^{\hat{L}_{\CC}\, t}  |\Omega \rangle_{\CC} = \,  _{\CC} \langle \SS |\, \hat{\Psi}\,   |\Omega \rangle_{\CC} ,
\end{align}
\end{subequations}
%%%%%%%%%%%%%%%%%%%%%%%%
where $\hat{L}_{\CC}$ is the Liouvillian of system in representation $\CC$, $ |\,\rho (0) \rangle_{\CC}$ is the initial state of system and $|\Omega \rangle_{\CC}$ is the steady state of system. $\hat{\Psi}$ is the vector of operators in terms of correlation function vector $\Psi$, for example, in our model $\hat{\Psi}$ equals to $(a^\dag_{k_1} a_{k_2},\, a^\dag_{-k_2} a_{-k_1},\, a_{-k_1} a_{k_2},\, a^\dag_{-k_2} a^\dag_{k_1})^T$. Substituting Eq.(\ref{RE}) into Eq.(\ref{Gt}), we obtain
%%%%%%%%%%%%%%%%%%%%%%%%
\begin{equation}
_{\CC} \langle \SS |\, \hat{\Psi}\, e^{\hat{L}_{\CC}\, t} \big( \, |\,\rho (0) \rangle_{\CC} - |\Omega \rangle_{\CC} \big) =\,  _{\CC} \langle \SS |\, \hat{\Psi}\, e^{\Gamma\, t} \big( \, |\,\rho (0) \rangle_{\CC} - |\Omega \rangle_{\CC} \big).
\end{equation}
%%%%%%%%%%%%%%%%%%%%%%%%
Comparing the two sides of the above equation, we have
%%%%%%%%%%%%%%%%%%%%%%%%
\begin{equation}
e^{\hat{L}_{\CC}\, t} \big( \, |\,\rho (0) \rangle_{\CC} - |\Omega \rangle_{\CC} \big) = e^{\Gamma\, t} \big( \, |\,\rho (0) \rangle_{\CC} - |\Omega \rangle_{\CC} \big),
\end{equation}
%%%%%%%%%%%%%%%%%%%%%%%%
and thus the eigenvalue $\Gamma$ of damping matrix $\X$ is also the  eigenvalue of Liouvillian $\hat{L}_{\CC}$.


\begin{thebibliography}{99}
\bibitem{Exp1}  Although the coefficient of $\hat{L}_{k=0}$ is $0.5$ in the equation $\hat{L}=0.5\hat{L}_{k=0}+\sum_{k=0^+}^{\pi^-} \hat{L}_k$, the rapidity for $k=0$ can be still described by $\lambda_{\pm} (0)$ due to the degeneracy as $\zeta^{'}_i(k=0)=\zeta^{'}_{5-i}(k=0)$ and $\zeta_i(k=0)=\zeta_{5-i}(k=0)$.
\end{thebibliography}
\end{document}